\documentclass[final,5p,times,twocolumn]{elsarticle}

\usepackage{aas_macros}
\usepackage{bm}
\usepackage{color}
\usepackage{listings}

\usepackage{amssymb}

\definecolor{mygreen}{rgb}{0,0.6,0}
\definecolor{myblue}{rgb}{0.3,0.5,1.0}
\definecolor{mygray}{rgb}{0.5,0.5,0.5}
\definecolor{mymauve}{rgb}{0.58,0,0.82}

\newcommand{\lib}
{HEATCODE} 

\graphicspath{{./}{plots/}}

\journal{Computer Physics Communications}

\begin{document}

\begin{frontmatter}



\title{Calculation of Stochastic Heating and Emissivity of Cosmic Dust Grains\\ with Optimization for the Intel Many Integrated Core Architecture}


\author{Troy A. Porter$^{1}$}
\ead{tporter@stanford.edu}
\author{Andrey E. Vladimirov$^{1, 2}$}
\ead{avladim@galprop.stanford.edu}

\address{$^1$ Hansen Experimental Physics Laboratory, Stanford University, 452 Lomita Mall, Stanford, CA 94305-4085, USA \\
$^2$ Colfax International, 750 Palomar Ave, Sunnyvale, CA 94085, USA
}

\begin{abstract}
Cosmic dust particles effectively attenuate starlight.
Their absorption of starlight produces emission spectra from the 
near- to far-infrared, which depends on the sizes and 
properties of the dust grains, and spectrum of the heating radiation field.
The near- to mid-infrared is dominated by the emissions by very small grains. 
Modeling the absorption of starlight by these particles is, however, 
computationally expensive and a significant bottleneck for self-consistent 
radiation transport codes treating the heating of dust by stars. 
In this paper, we summarize the formalism for computing the stochastic 
emissivity of cosmic dust, which was developed in earlier works, and 
present a new library \lib\ implementing this formalism for the calculation
for arbitrary grain properties and heating radiation fields.
Our library is highly optimized for general-purpose processors with 
multiple cores and vector instructions, 
with hierarchical memory cache structure.
The \lib\ library also efficiently runs on co-processor cards implementing 
the Intel Many Integrated Core (Intel MIC) architecture.
We discuss in detail the optimization steps that we took in order to 
optimize for the Intel MIC architecture, which also significantly benefited
the performance of the code on general-purpose processors, and provide 
code samples and performance benchmarks for each step.
The \lib\ library performance on a single Intel Xeon Phi coprocessor 
(Intel MIC architecture) is $\sim 2$ times a
general-purpose two-socket multicore processor system with 
approximately the same nominal power consumption.
The library supports heterogeneous calculations 
employing host processors simultaneously with multiple coprocessors, and can
be easily incorporated into existing radiation transport codes.

\end{abstract}


\end{frontmatter}


\section{Introduction}\label{sec:introduction}

Astrophysical objects can be difficult to study in the ultraviolet (UV) 
and optical
wavebands because of the effective attenuation of the visible light by 
cosmic dust particles.
To understand the broadband electromagnetic emissions from them 
therefore requires radiation transfer (RT) calculations that 
self-consistently 
account for the dust absorption and scattering of the UV/optical light, and 
subsequent
re-emission of the absorbed radiation at near infrared (NIR) to far infrared
(FIR) wavelengths.
These calculations can be time consuming, particularly for determining 
the dust re-emission spectrum because it depends on the sizes of the 
emitting grains and their associated properties.

Dust grains with sizes $\gtrsim 1$~micron typically attain a thermal 
equilibrium with the heating radiation field because the time-averaged 
vibrational energy of a grain is greater than the energy of the heating 
photons.
The emissivity for such dust particles can be straightforwardly obtained by 
balancing the absorption and re-emission rates.
However, for smaller grains attaining thermal equilibrium is often 
not the case because the absorption of a photon results in 
stochastic heating temperature spikes. 
Most of the energy of the grain is re-emitted in these events 
and no thermal equlibrium is attained.
Consequently,
stochastic grain heating cannot be modeled using methods that assume thermal 
equilibrium.

Methods for treating the stochastic heating of very small dust grains 
have been developed \citep[e.g.,][]{1985A&A...142L..19P,guhathakurta1989,1992A&A...266..501S,draine2001,2001ApJ...551..277M}.
However, in RT simulations of systems with a fine spatial 
voxelization, the calculation
of the stochastic grain emissivity for each cell can be prohibitively 
computationally expensive.
A typical approximation used to speed up these calculations involves
pre-calculating a table of emission spectra corresponding to different 
spectra of a parameterized synthetic heating radiation field 
\citep[e.g.,][]{2006MNRAS.372....2J,2013ascl.soft03030J,2012A&A...545A..39R}, 
and interpolating 
within the table on the basis of the incident total energy density of the heating 
radiation field in each simulation cell 
to obtain the stochastic heating emissivity.
While this approximation has in some cases been sufficient, it does not 
allow a fully self-consistent treatment for the small grain heating. 

The use of graphics cards to provide improvements in speed has been 
explored for Monte Carlo RT simulations by different authors 
\citep[e.g.,][]{2006MNRAS.372....2J,2012ApJ...751...27H}.
For the former, this usage was restricted to increasing the speed of 
the calculations for grains in thermal equilbrium with the heating radiation
field using 
Nvidia's CUDA architecture \citep{www-cuda}.
For the latter, the calculations for the stochastic and 
thermal equilibrium heating of grains were made employing CUDA.
However, a significant effort is often required to recode 
algorithms in the C-like CUDA language to exploit the hardware.
An attractive alternative is the recently released 
Intel Many Integrated Core (MIC) architecture 
\citep{www-mic}.
A significant advantage of 
the MIC architecture allows the same programming frameworks and 
optimization methods as those employed for multi-core general purpose 
processors\footnote{The term ``general purpose processor'' is used throughout this paper as a synonym for the terms CPU (Central Processing Unit), ``host'', and simply ``processor''. The term ``coprocessor'' is used interchangeably with ``Intel Xeon Phi coprocessor'' to denote an accelerator card implementing the Intel MIC architecture.}.
Therefore, the same code in a high-level programming language 
(C, C++ or Fortran) can be used to perform calculations on common general-purpose processors 
and coprocessor cards that implement the MIC architecture.

In this paper we describe the implementation and optimization of a common
method \citep{draine2001} for treating the stochastic heating of small 
dust grains that is based on a matrix formalism.
We provide details for a general-purpose processor target as well as the MIC
architecture.
Our goal was to make the stochastic grain heating emissivity calculation 
for an arbitrary radiation field efficient enough to make it feasible for fully
self-consistent calculations in simulations with high spatial 
voxelizations, while 
using essentially the same code for the general-purpose processors and the 
additional co-processor accelerators.
We show that significant effort must be invested into the optimization 
of the calculation to utilize the available compute power of the MIC 
architecture.
However, we find that most optimizations for the MIC architecture also 
improve the performance of the algorithm 
on the host system using general-purpose processors.
In contrast, with GPGPU\footnote{General-Purpose Graphics Processing Units}
-like architectures 
the accelerator code is generally distinct from the processor code,
and the architecture-level optimization of the application must be done
independently for the CPU code and for the GPGPU code.

The result of the work presented in this paper is a new 
astrophysical tool, the \lib\footnote{HEATCODE is an acronym for HEterogeneous Architecture library for sTochastic COsmic Dust Emissions}\ library, which efficiently computes the 
stochastic grain emissivity for arbitrary heating radiation fields both 
for general-purpose CPUs and co-processor cards implementing the Intel MIC 
architecture. 
In Section~\ref{sec:theory} we summarize the theoretical model for 
stochastic emissivity calculation.
Section~\ref{sec:performance} summarizes the performance improvement 
measures that were taken in order to optimize \lib\ for the multi-core and 
many-core architectures and demonstrates the performance on a 
benchmark server equipped with coprocessors featuring the 
Intel MIC architecture.
We summarize our findings in Section~\ref{sec:summary}.
\ref{sec:library} describes the user interface of the \lib\ library and 
outlines the structure of the code.
\lib\ can be obtained freely from the CPC Program Library.
In \ref{sec:optimization} the optimization steps are described in detail.

\section{Stochastic Heating of Very Small Dust Grains}\label{sec:theory}

Cosmic `dust' is a mixture of particles of different sizes and compositions; 
see the book by \citet{2011piim.book.....D} for an overview.
For a given grain species, e.g., silicates, graphitic grains, there is a 
distribution of sizes for the particles with the optical properties 
depending on the grain species.
The size distribution is, in general, unknown but parameterizations for 
various grain mixtures have been developed to explain features in the
extinction curves for the Milky Way and other 
galaxies, e.g., \cite{1977ApJ...217..425M,2001ApJ...548..296W,2004ApJS..152..211Z}.
HEATCODE can take the size distribution and optical properties as 
user-supplied inputs. 
However, for convenience we have included the 
size distributions from the authors referenced above, along with the grain 
absorption cross sections 
for silicate, graphite, and polycyclic aromatic hydrocarbons (PAHs) from 
B.\ Draine's 
website\footnote{http://www.astro.princeton.edu/\textasciitilde draine/dust/dust.diel.html}; 
our calculations below use the PAHs/graphite/silicate mixture and optical
properties from 
\citet{2001ApJ...554..778L} and the
size distributions for Milky Way dust from \citet{2001ApJ...548..296W}. 

For a given dust species, 
the heating radiation field is an input parameter supplied to the \lib\ 
library, 
and the spectrum of the re-emitted photons is the output of the calculation.
This is obtained by first calculating the absorption of 
stellar light and the spectrum of re-emitted light produced 
when transient excited states of the grains relax for each grain size.
Determining the re-emitted spectrum for each grain size 
involves two steps.
The temperature (or vibrational energy $E$) 
distribution of grains in response to the incident radiation spectrum is 
first computed.
We assume static equilibrium for this calculation, i.e., the stochastic 
grain heating is assumed to be exactly balanced by the grain relaxation with 
re-emission of photons.
This assumption is valid as long as the time scale of 
incident light variability is long compared to the relaxation times
of the transient excited grain states. 
Then, given the temperature distribution, the spectrum of 
re-emitted light for each grain size is calculated.
The output 
stochastic emission spectrum for the ensemble of grains is 
obtained by convolution of the individual grain size 
emission spectra with the grain size distribution.

\subsection{Calculation of the Temperature Distribution}

\citet{guhathakurta1989} proposed a matrix formalism for the treatment 
of the interaction between dust grains and photons that was further elaborated 
by \citet{draine2001}.
In their methodology, the temperature distribution of radiatively 
heated dust grains is described by ${\bm P}$, the ``state vector''.
The elements of this vector $P_i$ are the probabilities for the grains of the given size
to be in the energy bin $i$.
Energy bins span the range $\left[ E_\mathrm{min}, E_\mathrm{max}\right]$, and 
we assume $M+1$ narrow energy bins ($\Delta E \ll E$), so  $i=0, 1, \dots M$.

The
temperature distribution ${\bm P}$ is obtained by solving 
the system of linear algebraic equations that expresses balance between 
the population and depopulation rates of each energy level,

\begin{equation}
\sum_{j\ne i}T_{ij}P_j - \sum_{j\ne i}T_{ji}P_i=0 \quad \left(i = 0, 1, \dots, M\right),\\
\label{eq:steady-state}
\end{equation}
with the normalization condition
\begin{equation}
\sum\limits_{i=0}^{M} P_i = 1.
\label{eq:level-sum}
\end{equation}

\noindent
Here the matrix elements $T_{ij}$ are the probabilities per unit time for a 
grain in energy bin $j$ to transition to one of the states in bin $i$.

\citet{draine2001} describe the exact statistical treatment for the cooling transitions ($i<j$) as well as less computationally demanding thermal approximations for the corresponding cooling terms.
The thermal approach introduces the notion of grain temperature to approximate calculation of the downward transition rates.
The ``thermal discrete'' approximation allows discrete downward transitions to the same energy levels as the exact statistical treatment.
The ``thermal continuous'' approximation assumes that the only allowed spontaneous transition from an 
excited vibrational level $i$ is to the adjacent level $i-1$:

\begin{equation}
T_{ij} = 0,  \quad \mathrm{if} \quad i < j - 1.
\label{eq:adjacent}
\end{equation}
\noindent
This approximation has been shown by \citet{draine2001} to be accurate for 
grains comprised of as few as 30 atoms and allows for a fast solution method
compared to inverting the whole system of equations for the other methods.
The \lib\ library uses the ``thermal continuous'' approximation because it allows the computational cost of solving for the temperature distribution ${\bm P}$ 
to be considerably reduced while achieving reasonably accurate results.

The solution of the system of 
equations (\ref{eq:steady-state}), (\ref{eq:level-sum}) may seem 
straightforward.
However, the diagonal elements of the matrix corresponding to the system 
contain a difference of positive numbers with nearly equal values.
The solution can be unstable because of finite numerical precision.
\citet{guhathakurta1989} suggested that to minimize this effect on the 
solution, the system of equations (\ref{eq:steady-state}) can be 
transformed by adding to every equation all preceding equations. 
This opreation gives the following system:
\begin{equation}
\sum\limits_{j=0}^{f-1}B_{fj} P_j - T_{(f-1)f} P_{f} = 0
\quad(f=1,\dots,M),
\label{eq:bequation}
\end{equation}

\noindent
where
\begin{equation}
B_{fj} = \sum\limits_{k=f}^{M}T_{kj} \quad (f > j).
\label{eq:bmatrix}
\end{equation}

\noindent
Now the temperature distribution ${\bm P}$ can be determined by introducing 
a temporary vector ${\bm X}\equiv {\bm P}/P_0$ so that $X_0=1$.
Other values of ${\bm X}$ can be computed recursively:

\begin{equation}\label{eq:xf}
X_f = \frac{1}{T_{(f-1)f}} \sum\limits_{j=0}^{f-1}B_{fj} X_j \quad (f = 1, 2, \dots, M).
\label{eq:temperature}
\end{equation}

\noindent
Then the values of ${\bm P}$ are computed by re-normalization in order to 
satisfy Equation~(\ref{eq:level-sum}):
\begin{equation}
  P_{i} = \frac{X_i}{X_0 + X_1 + \dots + X_M} \quad (i=0,1,\dots,M).
\label{eq:renormalization}
\end{equation}

\subsection{Heating and Cooling Rates}

Heating transition rates are proportional to the incident light 
energy density $W(\lambda)$ at the wavelength corresponding to the transition 
energy $\Delta E_{ul}$,
and to the photon absorption cross section at this 
wavelength $\sigma(\lambda)$, which is a quantum mechanical property 
of the grains.
The heating rate is calculated as

\begin{equation}
T_{ul}  = W(\lambda)  \sigma(\lambda) \frac{\lambda^3 \Delta E_{ul}}{hc^2}  \quad \mathrm{for} \quad u>l.
\label{eq:heating}
\end{equation}

\noindent
The numerical wavelength grid on which $W(\lambda)$ and $\sigma(\lambda)$ are defined 
can map to an energy grid different from the grid used to represent the grain 
excitation levels.
Therefore, in the numerical solution, the 
quantity $W(\lambda)  \sigma(\lambda)\equiv \Omega(\lambda)$ in 
Equation~(\ref{eq:heating}) must be computed using interpolation:

\begin{equation}
\log \left[ \frac{\Omega(\lambda)}{\Omega(\lambda_{j-1})} \right] =
\frac{\log \left( \lambda / \lambda_{j-1} \right)}{\log \left(\lambda_j /\lambda_{j-1} \right)}
\log \left[ \frac{ \Omega(\lambda_{j}) } {  \Omega(\lambda_{j-1})  } \right].
\label{eq:interpolation}
\end{equation}

\noindent
Here $\lambda_{j-1}$ and $\lambda_{j}$ are, respectively, the lower and 
upper boundaries 
of the wavelength bins at which the energy density $W$ and 
cross section $\sigma$ are defined.

The rate of spontaneous relaxation to the adjacent level with photon 
emission is calculated as
\begin{equation}
T_{f-1,f} \approx \frac{1}{E_f - E_{f-1}}\frac{8\pi}{h^3 c^3}\int\limits_{0}^{E_f}\frac{E^3 \sigma(E) dE}{\exp(E/kT_f) - 1}, \; f>0.
\label{eq:cooling}
\end{equation}

\noindent
This rate does not depend on the incident radiation field, and therefore the 
convolution of the absorption cross section with the energy distribution 
function in Equation~(\ref{eq:cooling}) can be precomputed.

\subsection{Calculation of the Grain Emission Spectrum}

The emission spectrum for grains of size $a$ at wavelength $\nu = c/\lambda$ 
can be calculated using the following expression based on 
Equation~(56) in \cite{draine2001}:

\begin{equation}
\nu F_{a}(\nu) = \sigma(c/\nu) 
\sum\limits_{i=0}^{M} P_i (a) \Lambda(\nu, E_i),
\label{eq:emission}
\end{equation}
where 
\begin{equation}
\Lambda(\nu, E_i) = \left\{
\begin{array}{l}
0,\quad \mathrm{if} \quad E_i < h\nu, \\
\displaystyle\frac{2 h \nu^4}{c^2} \frac{P_i}{\exp(h\nu/kT_i) - 1} ,\quad \mathrm{otherwise}.
\end{array}
\right.
\label{eq:emission-lambda}
\end{equation}
The contribution of stimulated emission is neglected in this expression.

The combined emissivity from grains of all sizes is obtained by integration 
over the grain size distribution $Q(a)$:

\begin{equation}
\nu F(\nu) = \int\limits_{a_\mathrm{min}} ^{a_\mathrm{max}} \nu F_{a}(\nu) Q(a) da,
\label{eq:convolution}
\end{equation}
where $a_\mathrm{min}$ and $a_\mathrm{max}$ constrain the range of grain 
sizes considered in the model.
The maximum size $a_\mathrm{max}$ is chosen so that for $a>a_\mathrm{max}$, 
the radiative grain heating process is no longer stochastic, and can be 
treated in the thermal equilibrium approximation.

\subsubsection{Example}

\begin{figure}
\centering
\includegraphics[width=\columnwidth]{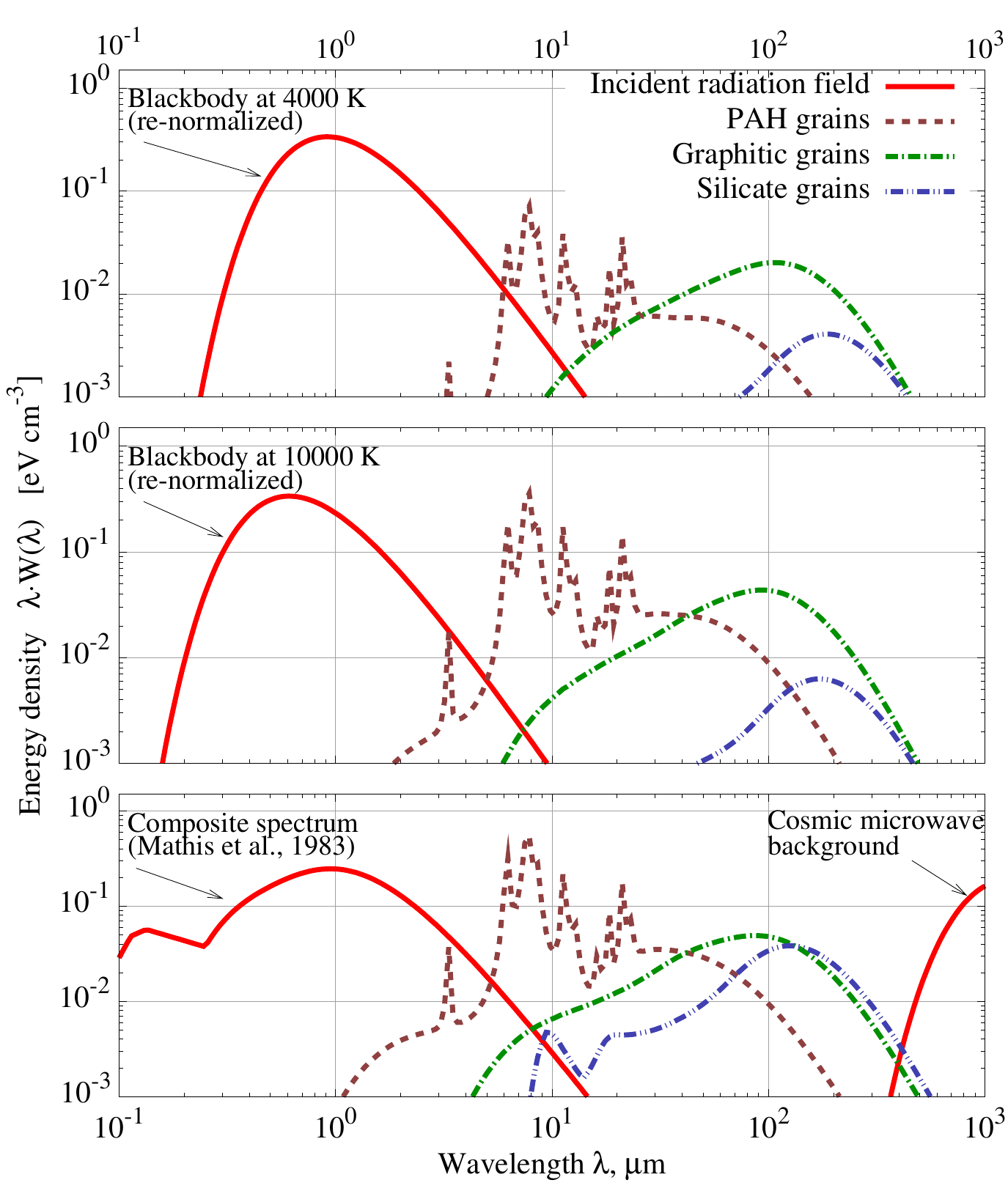}
\caption{Stochastic grain heating emissivities calculated by \lib\ for 
three incident light spectra. See text for details.\label{fig:intensities}}
\end{figure}

An example of the output of \lib\ is given in 
Figure~\ref{fig:intensities}, where the stochastic grain emissivity is 
computed for three heating radiation fields.
The heating radiation fields are plotted with solid red lines in each panel of 
the figure, and correspond to: (top) a blackbody spectrum at 4,000~K, 
(middle) a blackbody spectrum at 10,000~K, and (bottom) 
the interstellar radiation field (ISRF) from \citet{mathis1983}. 
The two blackbodies are normalized to the same energy density as the spectrum 
from \citet{mathis1983}, 0.46~eV~cm$^{-3}$. 
(For comparison with the heating radiation field energy densities, 
the Cosmic Microwave Background 2.735~K blackbody spectrum is shown in the
bottom panel.)
The calculated emissivity is converted to an equivalent energy 
density using the full solid angle and 
a gas column density $N_H=10^{22}$~cm$^{-2}$.
This conversion allows us to depict the incident radiation field and the 
dust-emitted light on the same plot for illustrative purposes.
Multiplication by the above mentioned gas column density
is an optically thin approximation of
the radiative transport throughout the Galaxy, assuming that 
\begin{enumerate}[a)]
\item the local radiation field is characteristic of the radiation field across the Galaxy, and
\item infrared light is not attenuated by absorption.
\end{enumerate}
Dashed and dash-dotted curves correspond to the emissions from 
the three grain models included 
in the library: graphitic, polycyclic aromatic hydrocarbons (PAH), and 
silicate grains.

\section{Performance Results}\label{sec:performance}

The \lib\ library can be used in self-consistent simulations 
that compute the stochastic heating and emissivity of small dust grains 
in every simulation cell using the in-cell heating radiation field.
Compared to an implementation of the algorithm relying solely on 
standard compiler optimizations and parallelization (e.g., the GNU C++ compiler
using OpenMP -- see Section~\ref{sec:benchmarks}) 
we obtain almost a factor of 100 times speed-up.
Achieving this performance is made possible by
\begin{enumerate}[a)]
\item Parallelization of calculations by distributing a set of incident radiation spectra across multiple processor cores;
\item Tuning various aspects of code performance for the general-purpose CPU architecture and for the Intel MIC architecture;
\item Relying on the Intel C++ compiler to perform automatic vectorization and other architecture-specific optimizations;
\item Performing the offload of the whole calculation or a part of it to Intel Xeon Phi coprocessors, if the computing system is enabled with them. Offload leads to additional speedup if the set of incident radiation fields is large enough.
\end{enumerate}
This section summarizes the optimizations applied to produce \lib\ and 
reports the benchmarks of performance.
For more detail on the optimization of stochastic heating and
emissivity calculation, refer to \ref{sec:optimization}.

\subsection{Performance Tuning}\label{sec:tuning}

\lib\ is optimized for general-purpose processors, and is able to run on 
Intel Xeon Phi coprocessors if they are available.
The performance tuning methods used in \lib\ apply to CPUs and Intel 
Xeon Phi coprocessors alike.
But, in this section we emphasize the optimization procedure for 
the coprocessors because of the novel nature of the architecture.

Intel Xeon Phi coprocessors that we used for benchmarks in the present paper
contain 60 independent cores (technical specifications stated hereafter are obtained from \cite{mic-sw-guide}).
These cores have a low clock frequency ($\sim$1~GHz) and four-way hyper-threading.
Each core relies on a dedicated vector arithmetics unit to maintain 
high performance with low power consumption.
The vector units support 512-bit vector registers for SIMD (Single Instruction Multiple Data) operations\footnote{For comparison,
the SSE registers supported by most general-purpose CPUs prior to the year 2011 are 128-bit long, 
and the AVX (Advanced Vector Extensions) registers supported by the Intel Sandy Bridge and AMD Bulldozer architectures are 256-bit long.}.
The  SIMD instruction set supported by Intel Xeon Phi coprocessors is called IMCI (Initial Many-Core Instructions).
IMCI includes floating-point multiplication, 
addition, fused multiply-add and division, some transcendental 
instructions, such as exponentials, logarithms and trigonometric functions,
and a range of bitwise logical operators, shuffle, gather/scatter memory accesses and bit-masked versions for most operations.

Onboard memory in Intel Xeon Phi coprocessors used for benchmarks in the present work
comes in the form of 8~GB of GDDR5 (Graphics Double Data Rate) with bandwidth 
capabilities typically exceeding that of the current generation CPU-based hosts.
The onboard memory is cached. 
The Level-2 caches of each core are linked with a ring interconnect.
This effectively merges local Level-2 caches of the cores into an aggregate coherent Level-2 cache of the coprocessor with non-uniform access and a typical access time of 11 cycles.
In addition, each core has a Level-1 cache with a typical access time of 1 cycle.
Coherency is maintained in all caches at the hardware level.
This means that after one core modifies data in a cache line, all other cores are guaranteed to see this modification when they access this cache line at a later time.

There are five major optimization methods used in \lib\ appropriate for 
the MIC architecture, as described below.

\subsubsection{Thread parallelism improvement}\label{sec:thread}
With $60 \times 4 = 240$ logical cores in the Intel Xeon Phi coprocessor, 
the performance-critical part of the application must have sufficiently large iteration space for work 
distribution, perform little synchronization between threads, and incur a small enough per-thread memory footprint.
In \lib, the first two issue are addressed by implementing thread parallelism in the following way:
\begin{enumerate}[a)]
\item the stochastic heating and emissivity for a single incident radiation field is calculated serially (i.e., by only one thread)
\item the library interface allows to call the emisivity calculation on a set of radiation fields corresponding to distinct voxels in the encompassing RT simulation
\item for processing a large number of radiation fields, multiple OpenMP threads are spawned, and each thread processes its share of radiation fields
\item dynamic loop scheduling in OpenMP is used to maintain load balance across threads
\end{enumerate}
As long as the number of input radiation fields supplied by the user is significantly greater than the number of threads (240),  all coprocessor cores are utilized.
This model of parallelism requires no synchronization between threads, because each radiation field is processed independently.
The third issue, the memory footprint of \lib, is addressed by optimizing the re-use of scratch space in each thread, so that, even with 240 threads, the application fits in memory.
This allows \lib\ to employ all of the logical cores, thus maximising the computational capacity of the co-processor.

\subsubsection{Vectorization refinements}

Within each thread, the calculation can be further parallelized through vectorization, i.e., by utilizing SIMD instructions of the computing platform.
Vector instructions apply the same arithmetical or logical operation to multiple numbers, thus multiplying the arithmetic performance by a factor equal to the width of the vector register.
The 512-bit vector registers in the MIC architecture can fit 16 single precision floating-point numbers or 8 double precision numbers, which corresponds to potential speedups of 16x and 8x, respectively, compared to scalar (non-SIMD) code.

We chose the most portable way to utilize vectorization in \lib: automatic vectorization by the compiler.
Automatic vectorization in the Intel C++ compiler is enabled by default at high optimization levels.
With this functionality, the compiler analyzes regular C++ loops, partitions them in blocks of 16 (or 8) iterations, and produces an assembly code that executes these blocks with SIMD instructions.
This occurs automatically and usually does not require the programmer's involvement.
However, we assisted the compiler by designing the C++ code of \lib\ with the following properties
\begin{enumerate}[a)]
  \item Data in automatically vectorized loops are accessed with unit stride (i.e., contiguously)

  \item The constructor of the  \lib\ data containers allocates memory structures so that the address of the first element of each array is a multiple of 64 (i.e., the data are 64-byte aligned), which enables faster loading of data into SIMD registers.
  
  \item The inner dimensions of multidimensional arrays are a multiple of 64 bytes, so that the first element of every row is 64-byte aligned.

  \item Most vector loops in \lib\ are constructed so that number of iterations is a multiple of 16 (the number of iterations was padded to a multiple of 16 when possible)

  \item Compiler hints in the form of vectorization pragmas were added to some vector loops in order to inform the compiler of data alignment, and to enable automatic vectorization in situations where the compiler could not detect whether vectorization is safe.
\end{enumerate}
With automatic vectorization, the C++ files of \lib\ can be compiled into exectutable code that runs with SIMD instructions on regular CPUs, as well as on Intel Xeon Phi coprocessors.

An alternative way to employ SIMD instructions would be through special functions called compiler intrinsics.
Intrinsics provide direct access to SIMD functionality, much like inline assembly.
However, sets of SIMD instructions available in CPUs and in the Intel MIC architecture are different.
This means that with intrinsics, the performance-critical portion of \lib\ would have to be written and optimized for every architecture independently.

\subsubsection{Scalar optimizations}

The thread-parallel, vectorized algorithm of the stochastic emissivity calculation in \lib\
can be compiled to perform in serial (i.e., single-threaded), scalar (i.e., non-vector) mode.
The more optimized that underlying scalar algorithm is, the better performance results are obtained when 
vectorization and thread parallelism are enabled.
Scalar optimizations employed in \lib\ are common for the many-core and multi-core 
architecture:
\begin{enumerate}[a)]

\item All floating-point calculations are performed in single precision; 
floating-point over/underflows are avoided by re-normalizing the data where necessary.
The algorithm with re-normalization is able to avoid floating-point exceptions as long as the input heating radiation
fields do not have integrated energy densities orders of magnitude larger
than the typical ISRF value of $\approx 1$~eV~cm$^{-3}$
Otherwise, the ratio between $x_f$ and $x_{f-1}$ in Equation~(\ref{eq:xf}) may be too large to handle in single precision

\item Optimized implementations for transcendental functions and strength reduction (substitution of multiple inexpensive operations for a single expensive operation) in arithmetic expressions are used. \ref{sec:step4} describes this in more detail.

\item In the constructor of the main library object, class \texttt{TransientHeatingModelXeonPhi}, 
precomputation of some quantities (e.g., the radiation field interpolation parameters) is performed.
The precomputed quantites are used during the evaluation of the stochastic heating and emissivity of dust (see \ref{sec:step4}).

\end{enumerate}

\subsubsection{Memory and cache traffic streamlining}

Main memory accesses in most computing architectures are more time-consuming 
than arithmetic operations.
This disparity is usually alleviated by storing a limited amount of data 
in a smaller, but faster memory known as cache.
The minimum amount of data that can be written to or read from a cache
is called a cache line.
Cache lines in Intel Xeon processors and Xeon Phi coprocessors are 64 contiguous bytes long.
In order to make the most out of caches, \lib\ is optimized in the following ways:
\begin{enumerate}[a)]
\item Data access in loops is made contiguous where possible. 
In addition, in order to avoid scattered memory access in the radiation field interpolation routine,
the interpolation parameters for each wavelength bin are precomputed and 
stored in a dense array (see \ref{sec:step3}).
These optimizations rely on the property that when one element of a cache line is accessed, the whole cache line with adjacent elements is served by the cache. 
Additionally, because of the predictability of streamlined access pattern, data can be automatically prefetched by the cache before it is even requested by the processor.
Contiguous memory access and the packing of data into dense arrays improve what is called  {\it data access locality in space}.

\item Nested loops are tiled (see \ref{sec:step5}), and consecutive loops of the same form are fused into one loop (see \ref{sec:step2}).
The effect of these optimizations is that when a portion of data is used multiple times in an algorithm, it is re-used as soon as possible after the previous usage, while it is still in cache.
These methods, nested loop tiling and loop fusion, improve what is known as {\it data access locality in time}.

\end{enumerate}

\subsubsection{Communication with the coprocessor}

\lib\ utilizes Intel Xeon Phi coprocessors using the offload programming model.
In this approach, the application runs on the host system (i.e., on the CPU), where it
initializes the physics data (cross sections, precomputed quantities related to the 
radiation field interpolation, Planck emissivity spectra, etc.) 
and accepts the input radiation fields from the user application.
With the data prepared on the host, \lib\ communicates with the coprocessor driver
and initiates offload (i.e., transfer) via the PCIe (Periperal Component Interconnect Express) bus of the necessary data and executable code to the coprocessor,
where the emissivity calculation functions are executed.
After that, the calculated emissivities are returned to the host system via the PCIe bus.

The design of \lib\ allows offloaded work to be distributed across multiple coprocessors, or across coprocessors and host CPUs.
This is done by splitting up the input array of incident radiation fields into chunks contatining multiple radiation fields, and offloading the next unprocessed chunk to the next available compute device.
In this process, the emissivity calculation function is offloaded to the coprocessor multiple times with different input data.
Because the physics data used by the calculation is the same in every offload, 
\lib\ optimizes the offload process by storing the physics data on the coprocessor between offloads.
In addition, to eliminate the overhead of memory allocation,
a persistent buffer is allocated 
on the coprocessor to hold the (input) radiation field array 
and the (output) stochatic heating emissivity array. 

The constructor of the main library object, class \texttt{TransientHeatingModelXeonPhi},
offloads all data related to the grain model or models to the 
coprocessor.
The desctructor of \texttt{TransientHeatingModelXeonPhi}  removes all 
persistent data and buffers from the coprocessor.

\subsection{Benchmarks}\label{sec:benchmarks}

For all benchmarks, we set up the runtime environment as described in 
\ref{sec:runtime}.
The size of the problem is chosen as \texttt{n}~$=10^4$ radiation fields 
for benchmarks using a single compute device, and \texttt{n}~$=10^5$ for 
heterogeneous calculations.
All radiation fields are set to the same spectrum for the benchmark, 
corresponding to the bottom panel of Figure~\ref{fig:intensities}.
The calculation time of all stochastic heating 
emissivities was measured 4 times, and the average of the last 3 is reported.
Multiple runs were used to control that the statistical accuracy of 
performance does not exceed the last reported significant figure.
Error bars are therefore omitted in all plots.

For the performance benchmarks, we used a server based on two Intel 
Xeon E5-2680 processors with 64~GB of DDR3 memory at 1,333~MHz.
This server hosts two Intel Xeon Phi coprocessors B1QS-5110P, each with 
60 cores and 8~GB of onboard GDDR5 memory.
The server runs a Linux operating system CentOS 6.4 on the host and 
the MPSS (Intel MIC Platform Software Stack) version 2.1.4982-15 on 
coprocessors.
Our tests were performed with one of the two compilers: Intel C++ compiler 
version 13.1.1 and GNU C++ compiler version 4.4.7. 
The former was used to benchmark the host, coprocessors, and the whole 
system in a heterogeneous calculation. 
The latter could be used only to benchmark the host processors, 
because the current GNU C++ compiler does not produce executable code
for the Intel MIC architecture.

\begin{figure}
\centering
\includegraphics[width=\columnwidth]{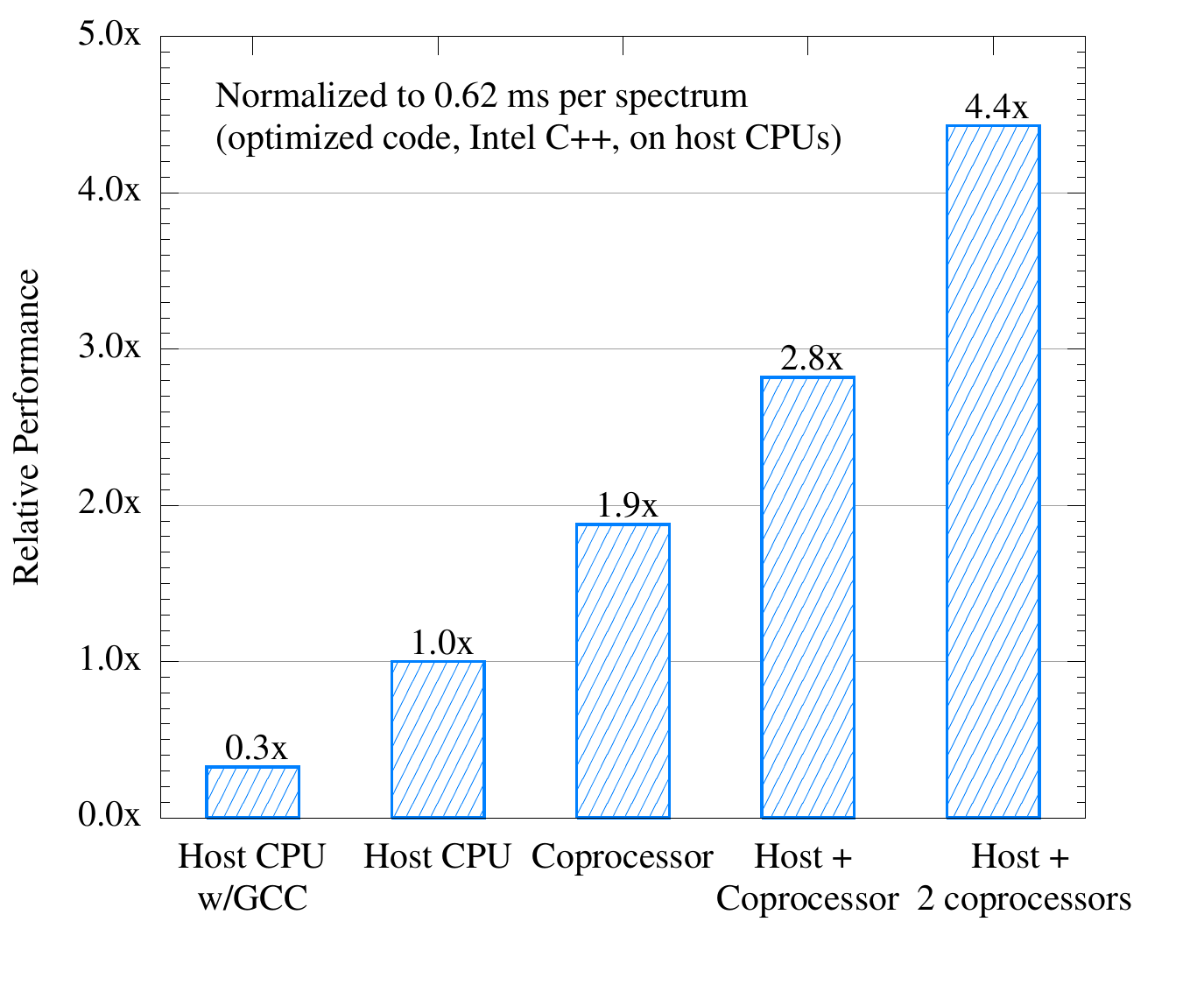}
\caption{Relative performance of the optimized version of \lib\ on the host, coprocessor, and in the heterogeneous mode.\label{fig:performance-opt}}
\end{figure}

The performance of the optimized stochastic emissivity calculation by 
class \texttt{TransientHeatingModelXeonPhi} in \lib\ is shown in 
Figure~\ref{fig:performance-opt}.
The performance values in the figure are normalized to the 
performance of the optimized code compiled with the Intel C++ compiler
and running on the host CPU.
Numerically, this normalization factor is $0.62$~ms per evaluation of the
stochastic emissivity function in a batch of \texttt{n}~$=10^4$ calculations.
This performance is a factor of 95x greater than the performance of the unoptimized code (hereafter referred to as the baseline performance).
The performance of optimized \lib\ compiled with GCC is more than 3 times lower 
than with the Intel C++ compiler on the same platform.

The difference between the performance of the calculation on a single 
Intel Xeon Phi coprocessor and on host system is a factor of 1.9x.
Note that this factor reflects the comparison of one Intel Xeon Phi 
coprocessor to a two-socket Intel Xeon processor.
The reason for comparing against two processor sockets is that two 
processors consume approximately the same power under load as one coprocessor.
The speedup of 1.9x is slightly lower than results of synthetic benchmarks 
SGEMM, SMP LINPACK and STREAM triad, which provide speedups 
of 2.9x, 2.6x and 2.2x, respectively \cite{benchmarks2013}. 
However, we find the 1.9x performance gain acceptable, considering two 
factors:
\begin{enumerate}[a)]
\item The workload of \lib\ has a latency-bound calculation phase with 
scattered memory accesses.
This type of workloads is better processed by many-core procesors than 
by the MIC architecture coprocessors.
\item The \lib\ is expressed completely in a high-level programming 
language with tuning via compiler pragmas.
This approach ensures cross-platform portability and forward compatibility. 
In contrast, some of the synthetic benchmarks mentioned above are tuned 
using low-level code with assembly instructions and/or intrinsic functions, 
which is not a portable solution.
\end{enumerate}

The last bar in Figure~\ref{fig:performance-opt} shows the performance of 
a heterogeneous calculation, which uses the host system simultaneously 
with two Intel Xeon Phi coprocessors. 
The performance gain of the heterogeneous calculation is 4.4x compared to 
the CPU-only calculation.
This is approximately 10\% lower than the sum of the raw performances of 
two coprocessors and host processors $1.9+1.9+1.0=4.8$x.
There is a performance loss caused by the unavoidable load imbalance 
across compute devices and the scheduling overhead of the 
heterogeneous calculation.
In our benchmarking 
we used a problem size of \texttt{n}=$10^5$ spectra; for a greater 
problem size, the performance loss due to scheduling would be lower.

\section{Summary and Discussion}\label{sec:summary}

We have presented \lib, a new library for the calculation of the 
stochastic heating and emissivity of cosmic dust grains for arbitrary 
incident radiation fields.
\lib\ can be run on general-purpose processors. 
In addition, it is optimized to also efficiently run on Intel Xeon Phi 
coprocessors and in heterogeneous systems with general-purpose 
processors and MIC architecture coprocessors.

Overall, the optimized \lib\ calculation running on our benchmarking system
and utilizing the host CPUs together with two Intel Xeon Phi coprocessors
processes 10$^5$ individual  stochastic emissivity spectra in 14 seconds,
on average $0.14$~ms per evaluation.
This is significantly faster than the unoptimized version (implemented in 
class \texttt{TransientHeatingModelUnoptimised}) compiled with GCC, which takes 
$6\cdot 10^{3}$ seconds for the same calculation ($60$~ms per evaluation).
An intended application of \lib\ is in conjunction with a RT code that is
used for calculating multiple instances of the observed local spectrum
of the interstellar radiation field and comparing each with observations, 
where each instance is evaluated for 
a set of Galactic structural parameters encoding, e.g., 
the locations of spiral arms, 
or the geometry of the Galactic bulge, over a parameter space.
The stochastic dust emissivity for every spatial grid cell must be calculated
for each realization of the parameters.
We estimate that $10^{5}$ RT calculations are required to explore
a representative space of 4~parameters, with $\sim10^5$ voxels for each 
realization of the Galaxy.
The single computing server with two Intel Xeon Phi coprocessors that we 
used for benchmarks spends $14\times 10^5$~s~$\approx 16$~days to compute 
the stochastic heating grain emissivities in such a calculation.
That is a reasonable duration for a research project, as opposed 
to $6\cdot 10^{3} \times 10^5$~s~$\approx 19$~years with the unoptimized 
code unable to use the coprocessor cards.

This performance improvement was made possible by using optimization 
practices that improve the performance of HPC applications on general-purpose 
processors and on the Intel MIC architecture.
An extended Appendix to this paper is dedicated to the discussion of these 
optimization practices.
As a supplement to this discussion, we provide in \ref{sec:optimization} nine code revisions 
corresponding to each optimization step.
This supplementary material in combination with the discussion in the 
text can be used to further improve the performance of the \lib\ library, 
and to transfer the methods practiced here to other applications with 
similar workloads.

The results discussed in \ref{sec:optimization} show that for almost all optimization steps, 
a performance increase is observed both in the host and in the coprocessor version of the library.
With the exception of minor fragments, the optimized C++ code is the same 
for all platforms and compilers.
Therefore, the common trend in all three curves in 
Figure~\ref{fig:performance} corroborates the claim that the optimization 
for the MIC architecture relies on the same principles as optimization 
for general-purpose processors  \citep{reinders2012}.
At the same time, the unoptimized code on the coprocessor performs a factor of 3x {\it worse} than the host system.
This observation supports the recommendation that applications must be 
optimized to fully exploit the general-purpose CPU capabilities before 
they can be efficiently run on the MIC architecture \citep{reinders2012}.

\section*{Acknowledgements}

The work on the \lib\ library is supported by NASA 
grants NNX12AO69G and NNX12AO73G.

\appendix

\section{The \lib\ Library Interface and Structure}\label{sec:library}

The \lib\ library is written in the C++ language and designed to function in the Linux environment. 
It can be compiled using the supplied Makefile, which supports the usage of the GNU Compiler Collection (GCC) and of the Intel C++ compiler.
The Intel compiler is necessary in order to utilize the MIC architecture support, which offloads parts of the calculation to Intel Xeon Phi coprocessors.
In addition, the Intel compiler produces a faster performing executable code than the GNU compiler, due to better support for automatic vectorization and single precision transcendental arithmetics.

The linking of \lib\ to a user application is done in the usual way by passing the corresponding object files to the linker. 
The compilation of an example application utilizing the \lib\ library is demonstrated in the Makefile.

\subsection{Interface}\label{sec:interface}

The interface to the library is demonstrated in the files \texttt{GrainModel.h} and \texttt{TransientHeatingModel.h}:
\begin{enumerate}[a)]
\item Class \texttt{GrainModel} is an abstract class which declares the data containers for the properties of dust interaction with light and size distribution of dust grains.
It must be overridden by concrete derived classes in order to implement specific grain models.
\lib\ contains derived classes \texttt{GraphiteGrainModel}, \texttt{PAHGrainModel} and \texttt{SilicateGrainModel} that implement the corresponding models of grains.
The constructor of \texttt{GrainModel} must be called with an input argument that defines the light wavelength grid used throughout the library.

\item Abstract class \texttt{TransientHeatingModel} instruments the interface and the computing engine (see Section~\ref{sec:theory}) that uses a \texttt{GrainModel} object to compute the stochastic emissivity of the corresponding type of grains heated by incident light spectrum.
\lib\ contains two concrete implementations of this class: \texttt{TransientHeatingModelUnoptimised}, which is an unoptimized version of the calculation, and \texttt{TransientHeatingModelXeonPhi}, which is optimized for heterogeneous calculations using general-purpose multi-core processors and Intel Xeon Phi coprocessors. 
The latter implementation is recommended for best performance even on systems not equipped with Intel Xeon Phi coprocessors.
The constructor of \texttt{TransientHeatingModel} must be called with two arguments that define the \texttt{GrainModel} class and the temperature grid used in the calculations.
The implementation \texttt{TransientHeatingModelXeonPhi} takes additional arguments that define the set of compute devices, which may be used for calculations.

\item The method \texttt{CalculateTransientEmissivity} of class \texttt{TransientHeatingModel} is the main interface function of the \lib\ library. Its signature is shown in Figure~\ref{code:cte}.
In this listing, classes \texttt{vector} and \texttt{valarray} are defined in the namespace \texttt{std} in the C++ Standard Library.
The input vector \texttt{rf} contains any number of incident radiation fields.
These radiation fields are direction-integrated light intensities measured in the units of eV~cm$^{-2}$~s$^{-1}$~m$^{-1}$ and defined on the wavelength grid used to initialize the class \texttt{GrainModel}.
The output vector \texttt{emissivity} upon the exit from the function contains the stochastic grain emissivities corresponding to each of the incident radiation fields.
The emissivity is the rate of energy emission multiplied by the wavelength. 
In the code, it is measured in the units of eV~s$^{-1}$~sr$^{-1}$~(H-atom)$^{-1}$ and defined on the same wavelength grid.

\begin{figure}[h]
\begin{lstlisting}[language=C++]
virtual void CalculateTransientEmissivity(
  const vector< const valarray<double>* > &rf, 
  const vector< valarray <double>* > &emissivity
  ) = 0;
\end{lstlisting}
\caption{Signature of the main interface function of \lib.\label{code:cte}}
\end{figure}

\end{enumerate}

\subsection{Dependencies}

\lib\ uses OpenMP in order to distribute calculations across multiple threads on many-core processors and Intel Xeon Phi coprocessors.
Therefore, applications using \lib\ must be compiled with the compiler argument \texttt{-fopenmp} with GCC or \texttt{-openmp} with Intel C++ compiler.

\subsection{Usage in Parallel Computing Platforms}

The function \texttt{CalculateTransientEmissivity} automatically distributes work across multiple CPU cores and, if requested and possible, performs offload of parts of the calculation to coprocessors.
Usually, there is is no need to call this function from a parallel region, because it exploits all parallelism in the hardware system available to it.
However, \texttt{CalculateTransientEmissivity} may be safely invoked from multiple OpenMP threads.
In this case, only one instance of the calculation proceeds at a time, because the calculation is executed inside a critical OpenMP barier.

\lib\ may be used in cluster environments if MPI processes operating on independent nodes call independent instances of the library.
However,  within \lib, distributed-memory parallelism with MPI is not supported.

When multiple instances of an application using \lib\ are running on a compute node with Intel Xeon Phi coprocessors, each instance will try to claim all available compute resources: CPU cores and coprocessors.
In cases when it is necessary to share these resources between several instances of the \lib\ library, the user must implement inter-process locking or otherwise modify \lib\ in order to avoid resource contention.

\section{Optimization of \lib}\label{sec:optimization}

In this section we discuss the optimizations that had to be made in order to improve the performance of the stochastic heating and emissivity calculation and utilize the Intel MIC architecture.
Most of these optimizations apply to all modern general-purpose processors and are not specific to the nature of the \lib\ algorithms.
We report our optimization work in detail in order to facilitate future development of \lib\ and of other applications relying on similar computational algorithms.
As of the writing of this work, the Intel MIC architecture is a young technology with only eight months of public exposure.
We believe that the sharing of best optimization practices at this time is essential to the evolution of the scientific developer community.

Optimization steps discussed below are revisions of the class \texttt{TransientHeatingModelXeonPhi}.
Most optimization steps address a specific category of hardware capabilities or algorithm properties; however, some optimization improve performance by affecting multiple issues. 
We emphasize such multi-issue optimizations in the discussion.
The explicit code of all optimization steps can be viewed, cross-examined and benchmarked by downloading the supplement to \lib\ in the CPC library.

The unoptimized code (we refer to it as ``optimization step~0'') is our own application
developed without the aim to optimize it for the Intel MIC architecture or to tune it for a specific C++ compiler.
The code of step~0 is not intentionally handicapped for the purpose of highlighting the optimization benefits.
It was designed as an architecture-oblivious, straightforward implementation of the matrix formalism of \citet{draine2001}.
The performance of the code in step~0 compiled with the GNU C++ compiler is hereafter referred to as the baseline performance.
Quantitatively, the baseline corresponds to an average of $59$~ms per stochastic heating and emissivity calculation (in a batch of $10^4$ spectra).

\begin{figure*}
\centering
\includegraphics[width=\textwidth]{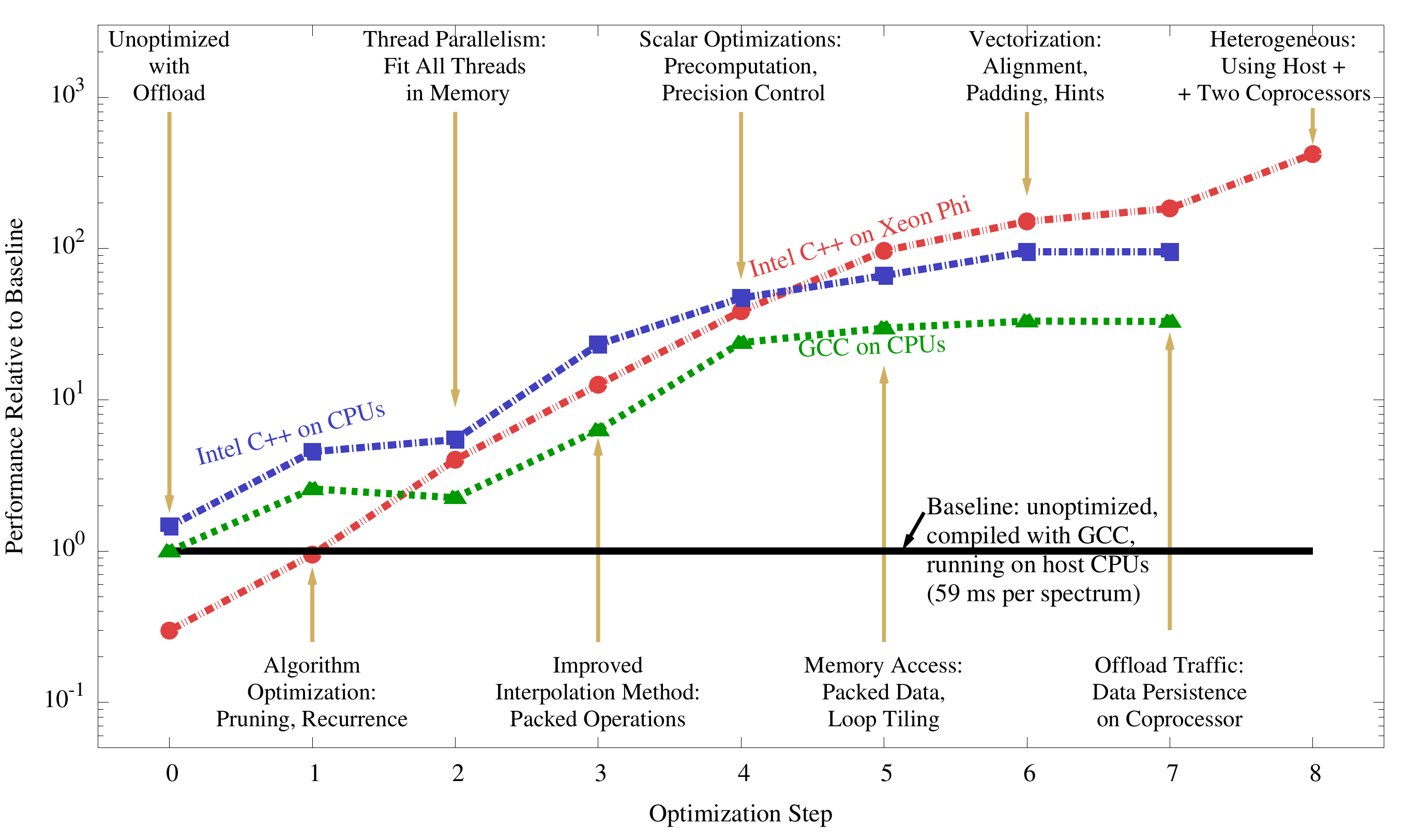}
\caption{Performance of \lib\ at optimization stages from the original implementation to the code highly optimized for heterogeneous multi- and many-core SIMD architectures.\label{fig:performance}}
\end{figure*}

Figure~\ref{fig:performance} shows the performance of \lib\ for all optimization steps with three curves. 
The green dashed line shows the performance of the code compiled with GCC on the host system utilizing all cores.
The blue dash-dotted line is the performance of the same code compiled with Intel C++ compiler, also on the host system.
The red dash-double-dot line is the performance on a single Intel Xeon Phi coprocessor (for steps~0 through~7), or on the whole system with the host and two coprocessors used concurrently (step~8).
All benchmark results shown in Figure~\ref{fig:performance} are measured relative to the baseline.

The performance of the optimized GCC version (step 7) is approximately 32x better than the baseline.
The Intel C++ compiler version on the host is approximately 95x faster than the baseline, and the code compiled with Intel C++ compiler on the coprocessor is 180x faster.
The heterogeneous system comprised of the host processors and two coprocessors performs 420x faster than the baseline version.

With GCC, the performance falls short of the corresponding performance with Intel C++ compiler.
Two factors contribute to this performance difference.
First, according to the vectorization report, GCC was not able to vectorize some of the loops in the code, including the calculation of the interpolated radiation field.
That calculation involves costly transcendental arithmetic operations and takes a significant fraction of the computation time.
Second, the Intel Math Library provides optimized exponential and logarithm functions with base 2, which are faster than their natural base counterparts, and \lib\ takes advantage of these optimized functions.
The math library used by GCC does also provides base 2 exponential and logarithm functions, but they are not faster than the natural base logarithm and exponential.
In order to compensate for that, when GCC is used for compiling \lib, the code falls back to using the natural base exponential and logarithmic functions.

In the rest of the Appendix, the specific measures taken in each optimization step are discussed and supplemented with fragments of C++ code and explanatory diagrams.

\subsection{Algorithm Optimizations: Pruning, Recurrence}\label{sec:step1}

Some applications can be improved by eliminating redundant operations from the algorithm.
In the context of computer algorithms, this procedure is known as ``pruning''.
Pruning is generally hardware-independent, but may require adopting a set of assumptions.
For calculations in \lib, two assumptions are made:
\begin{enumerate}[1)]
\item We assume that above a certain threshold grain size, grains become large enough so that the stochastic heating does not have to be computed.
Instead, adiabatic heating and emission formalism can be used, which is outside the scope of the \lib\ applicability.
In step~1, we optimize the code by precomputing the index \texttt{gIMax}, which corresponds to the threshold grain size.
Using this index, we restrict the bounds of corresponding loops of and reduce the amount of memory allocated for some storage containers.

\item The radiation field is assumed to be zero outside the wavelength range covered by the numerical wavelength grid.
Accordingly, the range of grain excitation energies can be constrained. 
We optimize the code by precomputing the index of the highest exited level for each grain size.
This index is represented by array \texttt{int fMax[gIMax]}.
Using \texttt{fMax}, we restrict the loop bounds in the respective matrix operations.

\end{enumerate}

Additionally, in step~1, we implement the pruning of grain sizes for which all heating terms calculated in the course of radiation field interpolation are zero.
This happens if the radiation field is non-zero only in a limited waveband, which does not interact with grains of a given size.
For such non-interacting grains, the temperature distribution is immediately obtained as $P_i=\delta_{i0}$, where $\delta_{ij}$ is the Kronecker delta. 
The pruning is done by counting and storing the number of non-zero heating terms in array \texttt{int nNonZero[gIMax]} and using this array in all functions to skip through trivial grain sizes.
This optimization does not involve any assumptions.

Another approach to algorithm optimization is the use of recurrence relations.
These relations express a quantity in a computed sequence via one or more preceding values in the sequence.
As \citet{guhathakurta1989} mention, the elements of matrix $B_{fj}$ defined by Equation~(\ref{eq:bmatrix}) can be computed using a recurrence relation, which in our notation is
\begin{equation}
B_{f,i} = 
\left\{
\begin{array}{ll}
B_{(f+1),i} + T_{f,i} & \mathrm{for} \quad i=0,1,\dots,M-1; \\
T_{f,i} & \mathrm{for} \quad i=M.
\end{array}
\right.
\label{eq:brecurrence}
\end{equation}

In optimization step~0 (previous), the calculation of $B_{ij}$ is performed explicitly (i.e., without recurrence) using the code shown in the left-hand side panel of Figure~\ref{code:bmatrix}.
In step~1 (currently discussed), we modify this code as shown in the right-hand side panel of Figure~\ref{code:bmatrix}.
This optimization significantly improves the performance by reducing the cost of the calculation of $B_{ij}$ from $O\left(M^3\right)$ to $O\left(M^2\right)$.
Figures~\ref{code:bmatrix}, variables \texttt{transientMatrix} and \texttt{bMatrix} are 3-dimensional arrays containing, respectively, matrices $T_{ij}$ and $B_{ij}$ for each grain size.
The variable \texttt{gI} is the index of the grain size, and \texttt{tempBins} is the size of the temperature grid corresponding to $(M+1)$ in Equations~(\ref{eq:bmatrix}) and (\ref{eq:brecurrence}).

\begin{figure*}

\begin{tabular}{cc}

\begin{minipage}{0.49\textwidth}
\begin{lstlisting}
for (int f = 0; f < tempBins; ++f){
	for (int i = 0; i < tempBins; ++i) {

		double sum = 0;
		for (int k = f; k < tempBins; ++k)
			sum += transientMatrix[gI*tempBins*tempBins + k*tempBins + i];
			
		bMatrix[gI*tempBins*tempBins + f*tempBins + i] = sum;
	}
\end{lstlisting}
\end{minipage}

&

\begin{minipage}{0.49\textwidth}
\begin{lstlisting}
double rSum[tempBins];
for (int i = 0; i < tempBins; i++)
	rSum[i] = 0;

for (int f = fMax[gI]; f >= 1; --f)
	for (int i = 0; i < f; ++i) {
		rSum[i] += transientMatrix[gI*tempBins*tempBins + f*tempBins + i];
		bMatrix[gI*tempBins*tempBins + f*tempBins + i] = rSum[i];
	}
\end{lstlisting}   
\end{minipage}

\end{tabular}

\caption{Left: unoptimized calculation of $B_{ij}$ for each grain size. Right: calculation of $B_{ij}$ for optimized by using a recurrence relation.\label{code:bmatrix}}
\end{figure*}

\subsection{Thread Parallelism: Inter-Procedural Loop Fusion for Reduced Memory Footprint}\label{sec:step2}

While the host system used in our tests has 32 logical cores, each of its two coprocessors has 240 logical cores.
As discussed in Section~\ref{sec:thread}, \lib\ employs thread parallelism by distributing across processor cores a large number of radiation fields for which the stochastic heating and emissivity must be computed. 
Intel Xeon Phi coprocessors used for testing have 8~GB of onboard memory, with some of it used for administrative purposes.
Each thread allocates scratch data in order to perform the calculation of stochastic heating and emissivity.
The greatest amount of memory is used for scratch arrays
which hold the results of intermediate calculations, and occupy up to $60$~MB per thread (the exact size depends on the grain model). 
On the coprocessor, each of the 240 threads allocates a private copy of all three arrays, resulting in a memory footprint in excess of $10$~GB.
Therefore, in order to run the calculation, we have to reduce the number of threads down to an amount that allows all scratch data to fit in memory, sacrificing the performance of unused cores.
In optimization step~2 (current), we resolve this situation as described below.

In each parallel thread spawned by the function \texttt{CalculateTransientEmissivity}, three arrays are used to temporarily hold scratch data of the calculation: \texttt{weightedRadiationField}, \texttt{transientMatrix} and \texttt{bMatrix}.
The size of each of these arrays is \texttt{gIMax}~$\times$~\texttt{tempBins}~$\times$~\texttt{tempBins}. 
The factor \texttt{gIMax}~$\gtrsim100$ occurs because the calculation stores one matrix of size \texttt{tempBins}~$\times$~\texttt{tempBins} for each grain size.
This is necessary because the unoptimized calculation is structured as a pipeline.
Three functions process the radiation field into the temperature distribution.
One function performs the radiation field interpolation (\ref{eq:interpolation}) and places it into the array \texttt{weightedRadiationField}.
The second uses the interpolated radiation fields to prepare matrices $T_{ij}$ and $B_{ij}$ stored in \texttt{transientMatrix} and \texttt{bMatrix}, respectively.
The third function uses \texttt{bMatrix} to solve, for each grain size, the system of equations (\ref{eq:bequation}) with these matrices.
The structure of the unoptimized calculation is shown in Figure~\ref{code:unfused}.

\begin{figure*}
\centering
\includegraphics[width=\textwidth]{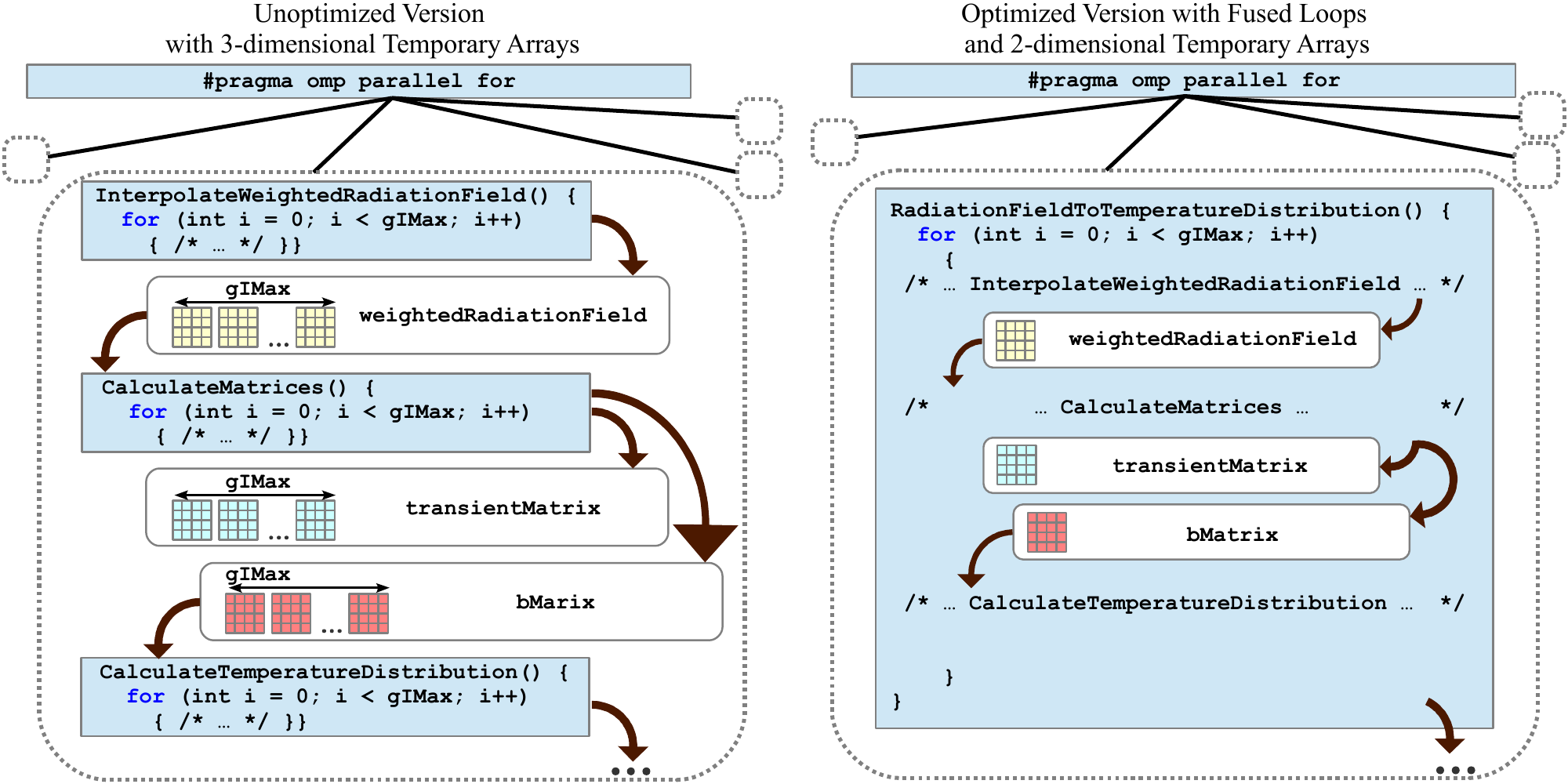}
\caption{Inter-procedural loop fusion reduces the memory footprint, which increases the usable number of OpenMP threads.\label{code:fusion}}
\end{figure*}

The optimization that we performe in step~2 involves loop fusion.
Each of the three pipeline functions contains a loop in the variable \texttt{gI}, which spans all grain sizes.
The calculation for each grain size is independent of all other grain sizes.
Therefore, if these three loops are fused into one, then, instead of a set of 3-dimensional arrays of size \texttt{gIMax}~$\times$~\texttt{tempBins}~$\times$~\texttt{tempBins},  a set of 2-dimensional arrays of size \texttt{tempBins}~$\times$~\texttt{tempBins} can be used.
We perform the optimization by moving the code of the three functions into a single function named \texttt{RadiationFieldToTemperatureDistribution} and fusing the loop in \texttt{gI}.
Naturally, we could once again partition the body of the new \texttt{gI} loop into three functions operating on only one index \texttt{gI}.
However, we chose to keep the code in one function, aiming for the best performance.
Argument passing and function calling could incur additional performance overhead.
The optimized structure of the code is listed in Figure~\ref{code:fused}.
A schematic view of this optimization is shown in Figure~\ref{code:fusion}.

This optimization does not significantly affect the performance of the host system based on general-purpose Intel processors.
Indeed, the host system with 64~GB of memory and 32 logical cores has sufficient memory to hold the scratch data for all 32 threads even in the unoptimized case.
However, the performance on Intel Xeon Phi coprocessors is improved by loop fusion by a factor of 4.
One reason for this is that with reduced memory footprint, the calculation is able to run as many OpenMP threads as there are logical cores on the coprocessor.
Another reason is that instead of an expensive allocation of large scratch arrays on the heap, threads perform quick allocation of small scratch arrays on the stack.
Dynamic allocation on the heap is particularly costly on the Intel MIC architecture, because this operation is inherently sequential.

\begin{figure*}[tp]
\begin{lstlisting}
void InterpolateWeightedRadiationField(/*output*/ float* weightedRadiationField, ...) {
	for (int gI = 0; gI < gMax; gI++) 
		{ /* ... calculating weightedRadiationField[gI*tempBins*tempBins + offset] */ }
}

void CalculateMatrices( /*input*/ const float* weightedRadiationField,  /*scratch*/ float* transientMatrix,  /*output*/ float* bMatrix, ...) {
	for (int gI = 0; gI < gMax; gI++) 
		{ /* ... calculating bMatrix[gI*tempBins*tempBins + offset] */ }
}

void CalculateTemperatureDistribution( /*input*/ const float* bMatrix, ...) {
	for (int gI = 0; gI < gMax; gI++) 
		{ /* ... using bMatrix[gI*tempBins*tempBins + offset] */ }
}

void CalculateTransientEmissivity(...) {
	/* Reducing the number of OpenMP threads to fit the problem in memory */
	const int reducedNThreads = max(240, availableMemory / memoryPerThread);
	omp_set_num_threads(reducedNThreads);
#pragma omp parallel for
	for (int r = 0; r < nRadiationFields; r++) {
		/* Allocation of temporary arrays in each thread */
		float* weightedRadiationField = (float*) malloc(sizeof(float)*gIMax*tempBins*tempBins);
		float* transientMatrix = (float*) malloc(sizeof(float)*gIMax*tempBins*tempBins);
		float* bMatrix = (float*) malloc(sizeof(float)*gIMax*tempBins*tempBins);

		/* Passing data across functions */
		InterpolateWeightedRadiationField(weightedRadiationField, ...); 
		CalculateMatrices(weightedRadiationField, transientMatrix, bMatrix, ...);
		CalculateTemperatureDistribution(bMatrix, ...);
		
		/* ... */
	}
}
\end{lstlisting}
\caption{Unoptimized structure of stochastic heating and emissivity calculation. The processing pipeline involves three distinct functions. The output of one function is the input of the subsequent function. In order to pass the data from one function to another, 3-dimensional temporary arrays are allocated on the stack, which requires too much memory, and the number of logical cores must be reduced.\label{code:unfused}}
\end{figure*}

\begin{figure*}[bp]
\begin{lstlisting}
void RadiationFieldToTemperatureDistribution(/* ... */) {
	/* 2-dimensional instead of 3-dimensional arrays
		Stack allocation here is faster than heap allocation in the unoptimized version. */
	float weightedRadiationField[tempBins*tempBins];
	float transientMatrix[tempBins*tempBins];
	float bMatrix[tempBins*tempBins];

	for (int gI = 0; gI < gIMax; gI++) {
		/* Here go the calculations previously performed in the body of the gI loop in 
			InterpolateWeightedRadiationFields(...)
			CalculateMatrices(...)
			CalculateTemperatureDistribution(...)
		*/
	}
}

void CalculateTransientEmissivity(...) {
#pragma omp parallel for
	for (int r = 0; r < nRadiationFields; r++) {
		RadiationFieldToTemperatureDistribution(/* ... */)
	}
}
\end{lstlisting}
\caption{Optimized structure of stochastic heating and emissivity calculation. The processing pipeline is placed in one function, and three loops in variable \texttt{gI} are fused into one. This allows smaller 2-dimensional temporary arrays. Now the problem fits into memory even when all logical cores are used. In addition, stack allocation of scratch arrays is possible, which is much faster than the heap allocation of large arrays.\label{code:fused}}
\end{figure*}

\subsection{Efficient Interpolation Method: Packing and Precomputing Scattered Memory Accesses}\label{sec:step3}

Optimization for multi- and many-core architectures usually requires the maintenance of data access locality.
Locality in this context means that the data in the main memory requested by cores should be requested from adjacent addresses (``locality in space'') and, if these data are used multiple times in an algorithm, they should be re-used as soon as possible (``locality in time'').
These properties of data access make the best use of hierarchical memory caches.

One of the operations performed in the stochastic heating calculation algorithm has an inherently scattered pattern of memory access.
This operation is the interpolation of the incident radiation field from the wavelength grid on which it is defined onto the grid on which the absorption cross section of the grains is discretized.
This pattern of memory access is inefficient because it involves memory access with poor locality in space.
We also argue below that it is inefficient for vectorization (i.e., the use of SIMD instructions).
The unoptimized interpolation algorithm is shown in Figure~\ref{code:interpolation}.

There are a number of issues in the unoptimized interpolation algorithm that prevent the multi- and many-core architecture from using its full potential:
\begin{enumerate}[1)]
\item The function \texttt{std::lower\_bound} is employed to find the radiation field wavelength bin interacting with the current grain wavelength.
This function uses the binary search algorithm, which incurs poorly predictable branches and scattered memory accesses to array \texttt{wavelength}.
Due to the combinatorial, non-vector nature of this algorithm, the Intel MIC architecture is used sub-optimally.

\item The data are read from arrays \texttt{radiationField} and \texttt{absorptionCrossSection} with a scattered pattern according to the index \texttt{j} found in the binary search. 
This pattern complicates effective cache use. 
It also prevents the vectorization of the transcendental expressions used to compute the interpolated value.

\item The data written to the matrix \texttt{weightedRadiationField} are re-used in the code when the matrix \texttt{transientMatrix} is computed, however, the calculations of these two matrices are performed in distinct loops.
This leads to sub-optimal data access locality in time.
By the time the calculation of \texttt{transientMatrix} begins, the bulk of \texttt{weightedRadiationField} is likely to have been evicted from processor caches into the main memory.
Loading \texttt{weightedRadiationField} from memory is the bottleneck for the calculation of \texttt{transientMatrix}.

\end{enumerate}

In step~3, we optimize the algorithm of radiation field interpolation and the computation of \texttt{transientMatrix} using several measures, as shown in Figure~\ref{code:interpolation-opt}:
\begin{enumerate}[a)]
\item First, we interchange the order of nesting in the traversal of the two wavelength grids.
Instead of traversing the grain wavelength grid first and for each set of indices (\texttt{f}, \texttt{i}) computing the index \texttt{j} in the radiation field grid, we made the optimized outer loop iterate through index \texttt{j} and compute the indices (\texttt{f}, \texttt{i}) for it.
This improves the pattern of memory access, because the index \texttt{f*tempBins + i} remains constant or monotonically increases with increasing \texttt{j}.
Even though now the calculation accesses \texttt{weightedInterpolationField} with a scattered pattern, the performance increases because the latency of scattered writes can be overlapped with computation, unlike the latency of scattered reads. 
The overlap occurs automatically thanks to out-of-order computation in Intel Xeon cores and due to hyper-threading in both Intel Xeon and Intel Xeon Phi cores.

\item The optimized code precomputes during the initialization phase the set of values (\texttt{f}, \texttt{i}) resonant with each wavelength bin \texttt{j}.
These values are stored in array \texttt{interpolationPatternIndex}. 
An auxiliary array \texttt{interpolationPatternCount} stores the number of grain wavelengths in the former array that resonate with \texttt{wavelength[j]}.
This completely eliminates a sorted array search, along with the poorly predictable branches caused by it, from the interpolation code.

\item The algebraic expression for the interpolated radiation field value is simplified, and most of the terms in it precomputed and stored in array \texttt{interpolationLogOffset}. 
This reduces the number of transcendental functions called for every interpolation event from 9 to an average value in the range $(1 \dots 2]$.
The specific number depends on how many grain wavelength bins interact with the given radiation field bin, on average.

\item The calculation of the elements of \texttt{transientMatrix} is moved from a separate loop into the interpolation loop. 
This improves the temporal locality of data access, because the values of \texttt{weightedInterpolationField} are still in the processor's registers when they are re-used.

\end{enumerate}

The optimizations discussed above increase the code performance thanks to improvements on multiple fronts.
Loop interchange streamlines the data access pattern, replacing scattered reads with scattered writes.
Precomputation of wavelength indices and interpolation expressions eliminates a combinatorial search and reduces the volume of transcendental arithmetic operations and non-local memory accesses.
Fusing the calculation of \texttt{weightedRadiationField} with its subsequent usage to compute \texttt{transientMatrix} improves the locality of data access in time.

\begin{figure*}
\centering
\includegraphics[width=\textwidth]{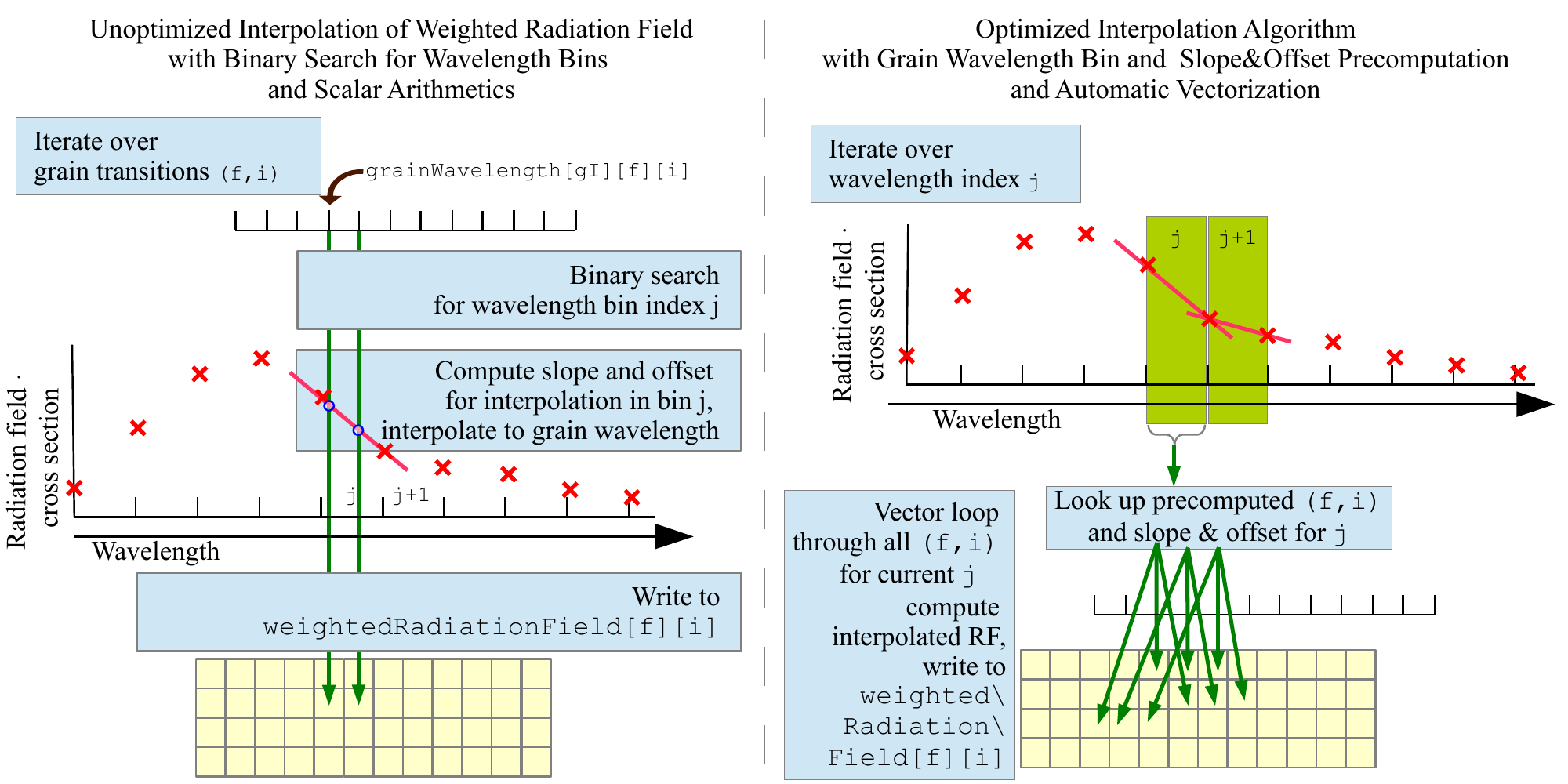}
\caption{Schematic interpolation algorithm before and after optimization.\label{fig:interpolation-schematic}}
\end{figure*}

A schematic diagram of the optimization of the interpolation algorithm is shown in Figure~\ref{fig:interpolation-schematic}.

\begin{figure*}
\begin{lstlisting}
for (int f = 1; f <= fMax[gI]; ++f) { /* Loop through all heating transitions for this grain size */
	for (int i = 0; i < f; ++i) {
		/* Wavelength at which cross section is defined */
		const double wl = grainWavelength[gI*tempBins*tempBins + f*tempBins + i];

		if (wl >= wavelength[0] && wl <= wavelength[wlBins-1]) { /* Guard against out of range results */

			const float* wlVal = std::lower_bound(&wavelength[0], &wavelength[wlBins-1], wl);
			const int j = wlVal - &wavelength[0]; /* Index of corresponding wavelength in radiation field grid */

			/* Two values of weighted radiation field to interpolate between: */
			const double lower = radiationField[j]*absorptionCrossSection[gI*wlBins + j];
			const double upper = radiationField[j-1]*absorptionCrossSection[gI*wlBins + j-1];

			if ((upper > 0) && (lower > 0)) {
				const double result = /* Linear interpolation in the log-log space */
					exp(log(lower) + (log(upper) - log(lower))*
					(log(wl) - log(wavelength[j-1]))/(log(wavelength[j]) - log(wavelength[j-1])));

				weightedRadiationField[gI*tempBins*tempBins + f*tempBins + i] = 
					result*(utl::m/utl::cm)*(utl::m/utl::cm); /* Conversion of units */
			}
		}
	}
}

/* transientMatrix is computed from weightedRadiationField in a separate loop */
for (int f = 1; f <= fMax[gI]; ++f)
	for (int i = 0; i < f; ++i)
		transientMatrix[f*tempBins + i] = weightedRadiationField[f*tempBins + i]*
			enthalpyDelta[gI*tempBins + f]*
			grainWavelength[gI*tempBins*tempBins + f*tempBins + i]*
			grainWavelength[gI*tempBins*tempBins + f*tempBins + i]*
			grainWavelength[gI*tempBins*tempBins + f*tempBins + i]*
			inversePlanckTimesSpeedOfLightSquared;

\end{lstlisting}
\caption{Unoptimized interpolation algorithm for the calculation of \texttt{weightedRadiationField} and the calculation of heating terms in \texttt{transientMatrix}.\label{code:interpolation}}
\end{figure*}

\begin{figure*}
\begin{lstlisting}
for (int j = 1; j < wlBins; j++) { /* Now outer loop traverses the radiation field wavelength grid */
	/* Two values of weighted radiation field to interpolate between: */
	const double lower = radiationField[j]*absorptionCrossSection[gI*wlBins + j];
	const double upper = radiationField[j-1]*absorptionCrossSection[gI*wlBins + j-1];

	/* How many grain wavelengths fall within the radiation field bin j (precomputed) */
	const int qCount = interpolationPatternCount[gI*wlBins + j];

	if ((upper > 0) && (lower > 0)) { /* Guard agains out of range results */ 
		const double dLogUpperLower = log(upper/lower); /* Same value for all bins */
		for (int c = 0; c < qCount; c++) {
			/* Index of the grain wavelength f*tempBins+i is precomputed during initialization */
			const int idx = interpolationPatternIndex[qCtr + c];

			/* Interpolation with precomputed log offset */
			weightedRadiationField[idx] = lower*exp(dLogUpperLower*interpolationLogOffset[qCtr + c]);

			/* Heating terms and unit conversion */                                                                       
			transientMatrix[idx] = weightedRadiationField[idx]*interpolationMatrixFactor[qCtr + c];
		}
	}
}
\end{lstlisting}
\caption{Optimized interpolation algorithm with improved spatial data locality (packed data) and precomputed expensive operations. The calculation of \texttt{transientMatrix} in the same loop as \texttt{weightedRadiationField} improves the data access locality in time.\label{code:interpolation-opt}}
\end{figure*}

\subsection{Scalar Optimizations: Precision Control, Precomputation}\label{sec:step4}

The preceding optimization steps (see \ref{sec:step1} through \ref{sec:step3}) eliminate algorithmically redundant operations, ensure scalable thread parallelism, facilitate automatic vectorization, and improve the memory access pattern.
The next step is optimization on the level of scalar operations by reducing the cost of arithmetics.

The unoptimized code uses mixed precision arithmetics. 
The usage of mixed precision is detrimental to performance for two reasons.
First, double precision operations take more time than their single precision counterparts.
Second, in vectorized mixed precision operations, additional type conversion and load/store operations are necessary, which may make mixed precision even less efficient than pure double precision.
In the unoptimized code, matrices \texttt{weightedRadiationField}, \texttt{transientMatrix} and \texttt{bMatrix} are single precision containers.
However, some intermediate arithmetic operations are performed in double precision in order to avoid floating-point underflow and overflow.

In optimization step~4 (currently discussed), we convert all floating-point variables and operations in the code to single precision.
In order to avoid overflow, we use scaling factors that decrease the magnitude of intermediate arithmetic results, so that the overflow does not occur.
These modifications are necessary in two locations in the code:
\begin{enumerate}[1)]

\item The calculation of the temperature distribution according to Equation~(\ref{eq:temperature}) can have overflow if the heating is strong.
In single precision, overflow occurs when for some $f>1$, the value $X_f$  is a factor of \texttt{FLT\_MAX}~$\approx 10^{38}$ greater than $X_0=1$.
We eliminate that overflow by checking that every new computed $X_f$ does not exceed a pre-determined threshold value (we used $10^{30}$ as the threshold).
If the value of $X_f$ exceeds the threshold, the whole vector ${\bm X}$ is multiplied by a factor $10^{-30}$.
This procedure, illustrated  in Figure~\ref{code:scaling}, affects the result only to machine precision, because eventually, the temperature distribution is calculated by re-normalization according to Equation~(\ref{eq:renormalization}).

\begin{figure}
\begin{lstlisting}
float sum = 0.0f;
for (int i = 0; i <= f - 1; ++i)      
	sum += bMatrix[f*tempBins + i]*x[i]; 

x[f] = sum*transientMatrix[(f-1)*tempBins + f];

if (x[f] > 1.0e30) {
	/* This code allows to use single precision for the entire solver. If the elements of x[] become too large, the vector is scaled down. */
	const float df = 1.0e-30;
	for (int i = 0; i <= f; i++) 
		x[i] *= df;
}
\end{lstlisting}
\caption{Optimized calculation of the temperature distribution with dynamic re-normalization avoids using double precision floating-point arithmetics.\label{code:scaling}}
\end{figure}

\item In the calculation of stochastic emissivity from the computed temperature distributions, out-of-range values occur because some physical quantities expressed in the SI units have extremely large or small absolute values. 
For example, photon absorption cross sections expressed in m$^2$ have values of order $10^{-20}$.
In the optimized code, we avoid using double precision by introducing the appropriate scaling factors for intermediate calculations with these quantities.

\end{enumerate}

In order to conform all quantities and functions in the code to the pure single precision implementation, two additional modifications are made:

\begin{enumerate}[1)]

\item All floating-point constants are explicitly marked as single precision. 
By default, floating-point constants in C++ expressions, such as ``\texttt{0}'', ``\texttt{1.0}'', etc., are treated as double precision numbers by the compiler. 
In order to explicitly declare them as single precision, the letter ``\texttt{f}'' should be appended: ``\texttt{0.0f}'', ``\texttt{1.0f}'', etc.
This avoids unnecessary type conversion in single precision expressions that use these constants.

\item Mathematical functions from the Math library optimized for single precision are used. 
That is, the double precision exponential function \texttt{exp()} is replaced with its single precision version \texttt{expf()}, and the logarithm \texttt{log()} with \texttt{logf()}.

\end{enumerate}

The exponential and logarithm transcendental functions can be further accelerated with the Intel Math library by using base-2 implementations \texttt{exp2f()} and \texttt{log2f()}. 
In the Intel Math library, these implementations are more efficient than the natural base versions \texttt{expf()} and \texttt{logf()}.
This is not true for the Math library used by the GNU compiler version 4.4.7, which we used.
Therefore, we implemented compiler-dependent code preprocessing  in order to use the most efficient exponential and logarithm functions available in the compiler used to build the application, as shown in Figure~\ref{code:exponentials}.
The preprocessor macro \texttt{\_\_INTEL\_COMPILER} is defined by the Intel C++ compiler; it is not defined by GCC.
This macro is also used to protect other syntax and functionality specific to the Intel C++ compiler, such as compiler vectorization hints (see below), and another automatically defined macro \texttt{\_\_INTEL\_OFFLOAD} is used to protect the offload pragmas that perform calculations on Intel Xeon Phi coprocessors.

\begin{figure}
\begin{lstlisting}
#ifdef __INTEL_COMPILER
/* In ICC
log2f is faster than log and faster than logf,
exp2f is faster than exp and faster than expf */
#define FASTLOG log2f
#define FASTEXP exp2f
#else
/* In GCC,
logf is faster than log and log2f,
expf is faster than exp and exp2f */
#define FASTLOG logf
#define FASTEXP expf
#endif

/* ... Later in the interpolation code: */
const float logUpperLower = FASTLOG(upper/lower);

/* Pre-computed pattern of access */
for (int c = 0; c < qCount; c++) {
	const int idx = interpolationPatternIndex[qCtr+c];
	/* Interpolation */
	weightedRadiationField[idx] = lower*FASTEXP(logUpperLower*interpolationLogOffset[qCtr+c]);
	/* ... */
}
\end{lstlisting}
\caption{Compiler-dependent choice of the fastest implementations of the logarithm and exponential functions for interpolation.\label{code:exponentials}}
\end{figure}

Finally, in order to speed up some arithmetic operations, we employ a procedure known as strength reduction.
This procedure algebraically transforms arithmetic expressions, aiming to replace long-latency arithmetic operations, such as division, with faster operations, such as multiplication.
The result may not be bitwise-equivalent to the original expression, however, this is acceptable in cases such as ours.
Some strength reduction had already been performed in the interpolation algorithm in optimization step~3 (see \ref{sec:step3}).
In the current optimization step, we replace some divisions with multiplications by the reciprocal number, as illustrated in Figure~\ref{code:strength}.
The code in this figure implements the re-normalization of the temperature distribution expressed by Equation~(\ref{eq:renormalization}).

\begin{figure}
\begin{lstlisting}
float pSum = 0.0f;
for (int i = 0; i < tempBins; i++)
	pSum += x[i];

const float invPSum = 1.0f/pSum;
for (int f = 0; f <= fMax[gI]; ++f)
	/* distribution[gI*tempBins + f] = x[f]/pSum; */
	distribution[gI*tempBins + f] = x[f]*invPSum;
\end{lstlisting}
\caption{Strength reduction example: replacing division with multiplication by the reciprocal value.\label{code:strength}}
\end{figure}

\subsection{Memory Access Optimization: Redundant Storage Elimination, Loop Interchange and Tiling}\label{sec:step5}

In optimization steps~2 and~3 (\ref{sec:step2} and \ref{sec:step3}) we already addressed the alleviation of some of the memory access issues by reducing the scratch storage size and improving the memory access locality in space and time.
In optimization step~5 (currently discussed), we continue memory access optimization and perform three additional code modifications.

\subsubsection{Sharing Memory Between Arrays}

First, we recognize that in the pipeline that processes the radiation field to compute the temperature distribution, three temporary arrays: \texttt{weightedRadiationField}, \texttt{transientMatrix} and \texttt{bMatrix} are used sequentially.
When the next array is filled, the data in the previous array are no longer needed.
This situation lends us the opportunity to further reduce the memory footprint and increase data locality by using a shared memory location for all three arrays.
We accordingly improve the code by allocating space for only one array and declaring another array as a pointer referring to the same memory location.
The array \texttt{weightedRadiationField} is eliminated, and only a temporary scalar variable is used in its place.
Some care must has to be taken in order to preserve automatic vectorization, because the compiler is wary of vector dependences in loops operating on arrays with overlapping memory locations.
The vectorization success is controlled using the compiler argument \texttt{-vec-report3} of the Intel C++ compiler in order to produce a detailed vectorization report.
This optimization and the construction of the modified loops is shown in Figure~\ref{code:overlap}.

\begin{figure}
\begin{lstlisting}
float bMatrix[tempBins*tempBins];
float* transientMatrix = &bMatrix[0];

/* ... Later in the interpolation algorithm: */
for (int c = 0; c < qCount; c++) {
	const int idx = interpolationPatternIndex[qCtr+c];
	/* Array weightedRadiationField is eliminated: */
	const float weightedRadiationField=lower*FASTEXP(logUpperLower*interpolationLogOffset[qCtr+c]);
	transientMatrix[idx] = weightedRadiationField*interpolationMatrixFactor[qCtr + c];
}

/* ... Later in the calculation of bMatrix: */
for (int f = fMax[gI]; f >= 1; --f)
	for (int i = 0; i < f; ++i) {
	/* bMatrix and transientMatrix share the same memory space. Previously used line
		rSum[i] += transientMatrix[f*tempBins + i];
	was modified in order to preserve efficient vectorization: */
		rSum[i] += bMatrix[f*tempBins + i];
		bMatrix[f*tempBins + i] = rSum[i];
	}
\end{lstlisting}
\caption{Improving memory traffic by sharing memory space between scratch arrays and loop fusion for \texttt{weightedRadiationField} and \texttt{transientMatrix} calculation.\label{code:overlap}}
\end{figure}

\subsubsection{Loop Interchange}

In the calculation of stochastic emissivity from the temperature distribution, expressed by Equations~(\ref{eq:emission}) and (\ref{eq:convolution}), nested loops are used to traverse the emissivity wavelengths, grain sizes and the temperature bins.
In the unoptimized calculation, the wavelengths are traversed first, followed by grain sizes, followed by the temperature bins in the inner nested loop.
We have found that permuting the outer two loops (i.e., traversing grain sizes first and wavelengths afterwards) improves the performance.
The unoptimized code (before permutation) and the optimized code (after) are shown in Figure~\ref{code:emissivity-permuted}.

The difference in memory access between unoptimized and optimized codes is the pattern of read access to arrays \texttt{distribution} and \texttt{planckDistribution}. 
In the unoptimized code, the latter array has better access locality than the former.
Indeed, for every iteration of the outer loop in variable \texttt{i}, only elements of the \texttt{j}-th row of \texttt{planckDistribution} are used, but the entirety of array \texttt{distribution} is read from memory.
In contrast, with permuted loops in the right-hand side panel of Figure~\ref{code:emissivity-permuted}, for every iteration of the outer loop in variable \texttt{j}, only the \texttt{i}-th row of \texttt{distribution} is used, whereas \texttt{planckDistribution} is read from front to back.
The permuted loop version works faster because when the calculation is distributed across multiple threads, each thread is operating on the same copy of \texttt{planckDistribution}, but the data in \texttt{distribution} are private to each thread.
In Intel Xeon processors, hyper-threading places two threads per physical core,  and in Intel Xeon Phi coprocessors, four threads per physical core.
Therefore, when the code operates on thread-private \texttt{distribution}, two (on the CPU) or four (on coprocessor) instances of this array must be fetched into the cache of each core. 
In contrast, for shared \texttt{planckDistribution}, only one copy per core must be fetched for all threads on that physical core to share.
This means that the application pays a greater penalty for cache misses on \texttt{distribution}, which is why the permuted loop version, with better locality of access to \texttt{distribution}, works faster.

\begin{figure*}[t]
\begin{tabular}{cc}

\begin{minipage}{0.49\textwidth}
\begin{lstlisting}
for (int i = 0; i < wlBins; ++i) { /* i first */

	float sum = 0.0f;
	for (int j = 0; j < gIMax; ++j) {
		const float gsw = grainSize[j]*grainSizeDistribution[j];
		const float crossSection = absorptionCrossSection[j*wlBins + i];
		const float product = gsw*crossSection*scalingFactor;
		
		float result = 0.0f;
		/* i-th row of planckDistribution[] is re-used for all j-iterations, whereas distribution[] is read front to back */
		for (int k = 0; k < tempBins; ++k)
			result += planckDistribution[i*tempBins + k]*distribution[j*tempBins + k];
		sum += result*product;
	}
	trans[i] = sum*wavelength[i]*unitsConversion;

}
\end{lstlisting}
\end{minipage}

&

\begin{minipage}{0.49\textwidth}
\begin{lstlisting}
for (int i = 0; i < wlBins; ++i)
	trans[i] = 0.0f;
for (int j = 0; j < gIMax; ++j) { /* j first */
	const float gsw = grainSize[j]*grainSizeDistribution[j];
	for (int i = 0; i < wlBins; ++i) {
		const float crossSection = absorptionCrossSection[j*wlBins + i];
		const float product = gsw*crossSection*scalingFactor;
		
		float result = 0.0f;
		/* j-th row of distribution[] is re-used for all i-iterations, whereas planckDistribution[] is read front to back */
		for (int k = 0; k < tempBins; ++k)
			result += planckDistribution[i*tempBins + k]*distribution[j*tempBins + k];
		trans[i] += result*product;
	}
}
for (int i = 0; i < wlBins; ++i)
	trans[i] *= wavelength[i]*unitsConversion;
\end{lstlisting}
\end{minipage}

\end{tabular}
\caption{Left: Unoptimized code for stochastic emissivity calculation from the temperature distribution. Right: same code optimized code by a permutation of the outer loops.\label{code:emissivity-permuted}}
\end{figure*}

\subsubsection{Loop Tiling}

Further improvement of cache performance in the calculation of stochastic emissivity may be achieved with the use of loop tiling.
Loop tiling transforms two nested loops in such a way that 
\begin{enumerate}[a)]
\item the outer loop variable is incremented with a stride greater than one (the stride is called the tile size), and
\item for every value of the inner loop variable, a fixed small number of outer loop iterations are performed to traverse the tile.
\end{enumerate}
This technique is often used in matrix operations with nested loops in order to increase the re-use of data fetched into processor registers or cache before these data are evicted.
In the optimized version in step~5, we employ double tiling, applying the technique to both the \texttt{i} and \texttt{j} loops.
Some empirical tuning is required in order to find the optimal tile size.
The resulting code (incomplete) is shown in Figure~\ref{code:emissivity-tiled}.

\begin{figure*}
\begin{lstlisting}
const int iTile = 4; /* Size of i-tiles */
const int jTile = 4; /* Size of j-tiles */
assert(wlBins % iTile == 0);

for (int jj = 0; jj < gIMax-(gIMax%jTile); jj += jTile) /* Striding through j-tiles */
	for (int ii = 0; ii < wlBins; ii += iTile) { /* Striding through i-tiles */
		float result[iTile*jTile];
		for (int c = 0; c < iTile*jTile; c++)
			result[c] = 0.0f;

		for (int k = 0; k < tempBins; ++k) /* Calculations are vectorized in the k-loop */
			for (int c = 0; c < iTile; c++) { /* Inner loop inside the i-tile */
				/* Loop inside the j-tile is unrolled: */
				result[(0)*iTile+c] += distribution[(jj+0)*tempBins+k]*planckDistribution[(ii+c)*tempBins+k];
				result[(1)*iTile+c] += distribution[(jj+1)*tempBins+k]*planckDistribution[(ii+c)*tempBins+k];
				result[(2)*iTile+c] += distribution[(jj+2)*tempBins+k]*planckDistribution[(ii+c)*tempBins+k];
				result[(3)*iTile+c] += distribution[(jj+3)*tempBins+k]*planckDistribution[(ii+c)*tempBins+k];
			}

		for (int b = 0; b < jTile; b++) { /* Collecting results for the tile */
			const float gsw = grainSize[jj+b]*grainSizeDistribution[jj+b];
			const float commonFactor = gsw*scalingFactor;
			for (int c = 0; c < iTile; c++)
				sum[ii+c] += result[b*iTile + c]*absorptionCrossSection[(jj+b)*wlBins + ii+c]*commonFactor;
		}
	}

for (int j = gIMax-(gIMax%jTile); j < gIMax; ++j)
  /* Finishing the non-tilable part of the j-loop (code not shown) */
\end{lstlisting}
\caption{Loop tiling improves the locality of memory access in the calculation of stochastic emissivity spectrum.\label{code:emissivity-tiled}}
\end{figure*}

The optimizations of loop permutation and tiling in this function lead to very significant performance improvement on the multi-core platform, and an even stronger improvement on the MIC architecture.
These methods are often required for memory traffic optimization in a large class of matrix calculations with a low arithmetic intensity.
Without loop tiling, such applications may be bottlenecked by memory access latency, nullifying the arithmetic and bandwidth advantages of specialized computing architectures such as Intel Xeon Phi coprocessors.
We have contributed the code sample and optimization techniques demonstrated in this section to \cite{colfax2013}, where this method is discussed in greater detail.

\subsection{Vectorization Refinements: Data Alignment, Loop Bound Padding, Compiler Hints}\label{sec:step6}

This section describes code modifications that optimize the automatic vectorization of calculations in \lib.

\subsubsection{Alignment}

A memory address is said to be aligned on an $B$-byte boundary when the numerical offet value of address is a multiple of $B$.
Most SIMD instruction sets require that when data are loaded to or from vector registers, the virtual memory address of the data must be aligned in a certain way. 
For example, the SSE2 (Streaming SIMD Extensions 2) vector instructions require data alignment on a 32-byte boundary.
The memory alignment requirement is relaxed in the AVX instruction set supported by the latest, as of this day, generation of general purpose processors.
However, the IMCI instruction set supported by Intel Xeon Phi coprocessors requires 64-byte alignment.

A block of memory can be aligned on the heap starting at an aligned address using the function \texttt{\_\_mm\_malloc()}, which is an intrinsic of the Intel C++ compiler available in \texttt{malloc.h}.
In \lib, for compatibility with other compilers, we implement a similar function that builds on the standard \texttt{malloc()} operation to allocate 64-byte aligned arrays.
For the alignment of stack arrays, the Intel C++ compiler supports the corresponding attribute specifier.
In optimization step~6 (currently discussed), we apply these methods to align the first element of arrays participating in vectorized loops, as illustrated in Figure~\ref{code:alignment}.

In the case of multi-dimensional arrays, the alignment of the first element array on a 64-byte boundary may not be suffucient.
Indeed, if the length of the inner dimension of this array is not a multiple of 64 bytes, then the first element in the second row (and subsequent rows) is not aligned.
Therefore, in order to improve vectorization in nested loops operating on multi-dimensional arrays, the inner dimension must be a multiple of the alignment value.
We verify that this condition is satisfied using an \texttt{assert()} check; however, we could also pad the row length to a multiple of 64 bytes for generality.

\begin{figure}
\begin{lstlisting}
#include <malloc.h>
#define ALIGN_BYTES 64
#define FLOATS_IN_ALIGN_BYTES 16

/* 64-byte alignment of an array on the heap */
planckDistributionF = (float*) _mm_malloc(sizeof(float)*wlBins*tempBins, ALIGN_BYTES);
_mm_free(planckDistributionF);

/* 64-byte alignment of an array on the stack */
float bMatrix[tempBins*tempBins]
	__attribute__((aligned(ALIGN_BYTES));
	
/* In order for each row of the above arrays to be aligned, the following must be satisfied: */
assert(wlBins % FLOATS_IN_ALIGN_BYTES == 0);
assert(tempBins % FLOATS_IN_ALIGN_BYTES == 0);

\end{lstlisting}
\caption{Alignment of arrays on the heap and on the stack.\label{code:alignment}}
\end{figure}

\subsubsection{Padding of Loops}

The Intel C++ compiler is able to vectorize loops even if the iteration count in the loop is not a multiple of the SIMD vector length.
For instance, in Intel Xeon Phi coprocessors, a 512-bit SIMD vector packs sixteen single precision floating-point numbers.
A loop in which the number iterations \texttt{N} is determined at runtime is not guaranteed to have a multiple of sixteen iterations.
In these cases, the compiler produces executable code for a variety of loop lengths \texttt{N}.
Depending on the runtime value of \texttt{N} and memory alignment situation, the compiler may perform \texttt{N} modulo $16$ iterations at the beginning or at the end of the loop in a non-vector form, and complete the rest using SIMD instructions.

Despite the compiler's flexibility,
it is often beneficial for performance to ensure that the loop count is a multiple of 16.
We have found that the \lib\ calculations that involve the quasi-triangular matrix $B_{ij}$ can be accelerated by changing the inner loop bounds to a multiple of 16.
Figure~\ref{code:bounds} shows examples of code modifications that demonstrated the best performance.
In optimization step~6, we increase the inner loop bounds either to the nearest multiple of 16.
This increases the number of floating-point operations, but significantly decreases the calculation time due to uninterrupted pattern of vectorization.
The change in the nested loop pattern expressed by the first example in Figure~\ref{code:bounds} is illustrated in Figure~\ref{fig:vecpattern}.

\begin{figure}
\begin{lstlisting}
/* In the calculation of bMatrix... */
for (int f = fMax[gI]; f >= 1; --f) {
	/* Computing iMax such that (iMax - 1)%0 == 0, and f-1 <= iMax <= tempBins-1 */
	const int uB = (f-1)+(16-(f-1)%16)-1;
	const int iMax = (uB<=tempBins-1?uB:tempBins-1);

	/* Unoptimized: "for (int i=0; i<f; ++i)" */
	for (int i = 0; i <= iMax; ++i) {
		rSum[i] += bMatrix[f*tempBins + i];
		bMatrix[f*tempBins + i] = rSum[i];
	}
}

/* Later, calc. of temperature distribution: */
x[0] = 1.0f;
for (int f = 1; f <= fMax[gI]; ++f) {
	if (rTransientMatrixOverDiagonal[f] > 0.0f) {
		float sum = 0.0f;
		/* Unoptimized: "for (int i=0; i<=f-1; ++i)" */
		for (int i = 0; i < tempBins; ++i)
			sum += bMatrix[f*tempBins + i]*x[i];
		x[f] = sum*rTransientMatrixOverDiagonal[f];
		/* ... */
	}
}
\end{lstlisting}
\caption{Padding loop count to a multiple of 16 to improve the efficiency of automatically vectorized loops.\label{code:bounds}}
\end{figure}

\begin{figure*}
\centering
\includegraphics[width=\textwidth]{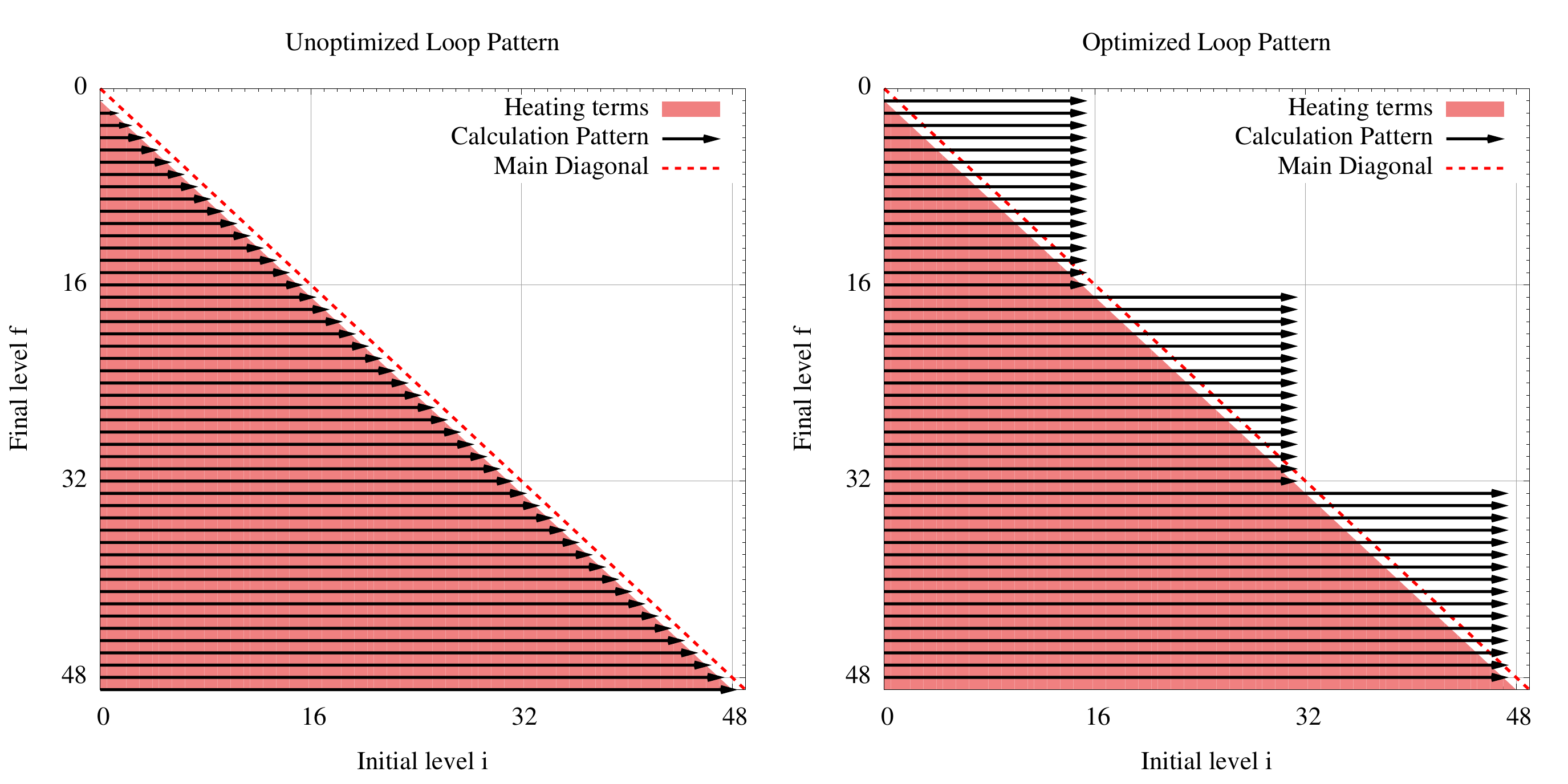}
\caption{Pattern of nested loops in \texttt{f} and \texttt{i} in the first example in Figure~\ref{code:bounds} before and after optimization. 
The optimized loop pattern always has a multiple of 16 iterations in the inner vectorized loop, which is beneficial for performance.\label{fig:vecpattern}}
\end{figure*}

In order to avoid overwriting the cooling terms $B_{(f-1)f}$ in this process, we store these terms in a separate matrix \texttt{rTransientMatrixOverDiagonal} instead of storing them at their respective locations in \texttt{bMatrix}.

Additionally, we modify the calculation of the pre-computed interpolation pattern (see Figure~\ref{code:interpolation-opt}) so that \texttt{interpolationCount[i]} is a multiple of 16.
This achieves two results.
\begin{enumerate}[a)]
\item The number of iterations of the loop in variable \texttt{c}  in Figure~\ref{code:interpolation-opt} is a multiple of 16, and
\item if \texttt{interpolationLogOffset[0]} is aligned on a 64-byte boundary, then in all instances of the \texttt{c}-loop, the element \texttt{interpolationLogOffset[qCtr+0]} is also 64-byte aligned. 
\end{enumerate}
Note that the GNU compiler was not able to vectorize this loop. The padding of \texttt{interpolationCount} to a multiple of 16  degrades the performance with GCC, because it increases the number of loop iterations. 
In order to avoid performance degradation, we restrict the padding of this array to the Intel compiler compilation by protecting the corresponding code with a check for the preprocessor macro \texttt{\_\_INTEL\_COMPILER}.

\subsubsection{Compiler Hints}

Even if the programmer ensures that every instance of a vector loop operates on aligned data, the compiler still implements runtime checks for alignment in order to maintain error-free execution.
However, it is possible to instruct the compiler to drop these checks and assume the data aligned, which accelerates the program execution.
This is done by including the hint \texttt{\#pragma vector aligned} before the vectorized loop.
If the data are not aligned at runtime, the use of this pragma results in a segmentation fault.

If the runtime length of some loops is known to the programmer, but cannot be determined by the compiler, \texttt{\#pragma loop count} may accelerate execution.
Using the loop count suggested by the developer, the compiler more efficiently chooses the vectorization strategy.

Another useful compiler hint is \texttt{\#pragma simd}, which makes it mandatory for the compiler to vectorize the loop following the pragma.
This is particularly helpful in combination with loop tiling (see \ref{sec:step5}), where the vectorized loop is not the inner loop.

The hint \texttt{\#pragma vector nontemporal} can be used in order to perform vector write operations in a loop directly into the main memory, bypassing the cache.
This is useful when the output data of a vector loop are not used until much later in the code, and there is no reason to store them in a cache.
Even though nontemporal write instructions incur eviction stalls immediately, the overall application performance may be improved by using \texttt{\#pragma vector nontemporal}, because caches are not contaminated with temporary data.

In optimization step~6, we supplement the code, where appropriate, with the compiler hints described above.
Figure~\ref{code:hints} demonstrates some of the loops tuned with the vectorization pragmas.
These pragmas are specific to the Intel C++ compiler, and GCC does not recognize them.
We protect the vectorization hints with a check for the preprocessor macro \texttt{\_\_INTEL\_COMPILER} in order to avoid warnings with GCC.

\begin{figure}
\begin{lstlisting}
/* Calculation of bMatrix with loop bound padding */
const int uB = (f - 1) + (16-(f-1)%16) - 1;
const int iMax = (uB<=tempBins-1?uB:tempBins-1);

#ifdef __INTEL_COMPILER
#pragma vector aligned /* No alignment check */
#pragma loop count min(16) /* Expected loop count */
#endif
for (int i = 0; i <= iMax; ++i) {
	rSum[i] += bMatrix[f*tempBins + i];
	bMatrix[f*tempBins + i] = rSum[i];
}

/* Later: calculation of emissivity, loop tiling */
#ifdef __INTEL_COMPILER
#pragma vector aligned /* No alignment check */
#pragma simd /* Vectorize outer loop */
#endif
for (int k = 0; k < tempBins; ++k)
	for (int c = 0; c < iTile; c++) {
		result[(0)*iTile+c]+=distribution[(jj+0)*tempBins+k]*planckDistribution[(ii+c)*tempBins+k];
		result[(1)*iTile+c]+=distribution[(jj+1)*tempBins+k]*planckDistribution[(ii+c)*tempBins+k];
		/* ... */
	}
\end{lstlisting}
\caption{Pragma hints indicate to the compiler how to best choose the automatic vectorization strategy for a loop.\label{code:hints}}
\end{figure}

\subsection{Offload Traffic Reduction: Data Persistence and Memory Retention on the Coprocessor}\label{sec:step7}

In optimization steps~1 through~6 (\ref{sec:step1} through \ref{sec:step6}) we focused on optimizing the calculations, whether they are running on the host system or on the coprocessors.
As a result, the execution time of the optimized code has decreased enough that the set-up time of the calculations has become significant.
In this optimization step, we reduce the set-up time for calculations offloaded a coprocessor.

\lib\ uses Intel Xeon Phi coprocessors in the explicit offload mode.
In this mode, before a function is executed on the coprocessor, all the data that the function accesses must be transferred across the PCIe bus from the host to the coprocessor.
Which data structures are sent, and in what direction, is controlled by the clauses of \texttt{\#pragma offload}, which initiates the offload.
By default, for each array transferred to the coprocessor, memory is allocated on the coprocessor, the data are copied across the PCIe bus, and then the calculation is run.
At the end of the offload region, the output data are sent back to the host, and memory on the coprocessor is deallocated.
However, this behavior may be modified to retain the data and/or memory on the coprocessor between offloads.
This may result in a significant decrease in the set-up time for two reasons.
\begin{enumerate}[a)]
\item The rate of data transfer is limited by the PCIe bandwidth, which, for large arrays, reaches 6 to 7~GB~s$^{-1}$.
\item However, an even more severe limitation is the rate of memory allocation on the coprocessor. 
Dynamic memory allocation is an inherently sequential operation, and the low-frequency cores of the coprocessor have an effective allocation rate of only around 0.5~GB~s$^{-1}$ (see \cite{colfax2013}).
\end{enumerate}

Because all calls to the stochastic heating and emissivity calculation use the same common data, such as grain absorption cross sections, precomputed interpolation pattern, grids, and other, it makes sense to retain these data between offloads in order to reduce the set-up time.
In order to retain the memory allocated for an array, the clause \texttt{free\_if(0)} is used, and in order to avoid duplicate memory allocation, \texttt{alloc\_if(0)} is placed in the list of clauses of \texttt{\#pragma offload}.
The copying of data can be avoided by specifying a zero length of the offloaded array, i.e., \texttt{length(0)} if the data had already been transferred to the coprocessor.

The situation is somewhat complicated by the fact that the encompassing ISRF calculation may use multiple grain models.
However, the offload runtime library is able to deal with multiple instances of arrays with the same name persistently stored on the coprocessor.
It manages persistent data by storing the mapping between host pointers to all offloaded arrays and the respective persistent arrays on the coprocessor.

Another offload-related optimization is the retention of memory for the array of incident radiation fields and the computed stochastic emissivities.
These data are not the same in all calls to \texttt{CalculateTransientEmissivityXeonPhi}, because the user or the encompassing ISRF calculation is expected to call the stochastic heating and emissivity calculation for different sets of radiation fields.
However, as mentioned earlier, the effective rate of memory allocation on coprocessors is an order of magnitude lower than the data transfer bandwidth.
Therefore, it is still beneficial to retain the memory allocated for the incident radiation fields and emissivities, even if the data have to be transferred and copied into this memory in every offload instance.
In order to allow arbitrary number of radiation fields to be passed to the function to \texttt{CalculateTransientEmissivityXeonPhi} (only limited by the memory capacity of the coprocessor), we implement memory management for the radiation field container array with dynamic expansion of the containers when necessary.

\subsection{Heterogeneous Computing: Concurrent Usage of Multiple Compute Devices}\label{sec:step8}

We have optimized all aspects of \lib\ to achieve good performance on the host multi-core system and on a single coprocessor.
The performance on a single coprocessor (see Section~\ref{sec:benchmarks}) is 1.9x greater than the performance on two host processors.
This number, given approximately equal power consumption in the host and in the coprocessor, reflects the performance per watt improvement.
However, for practical applications, we are interested in completing a calculation in the shortest possible time, which requires using all system resources in tandem.
That said, the application must be able to utilize the host simultaneously with the coprocessor or multiple coprocessors if they are present.
This functionality is extremely important because hardware system configurations available today can host up to eight Intel Xeon Phi coprocessors in a single host server.
We describe the implementation of heterogeneous and multi-coprocessor \lib\ calculations in this section.

\begin{figure*}
\centering
\includegraphics[width=\textwidth]{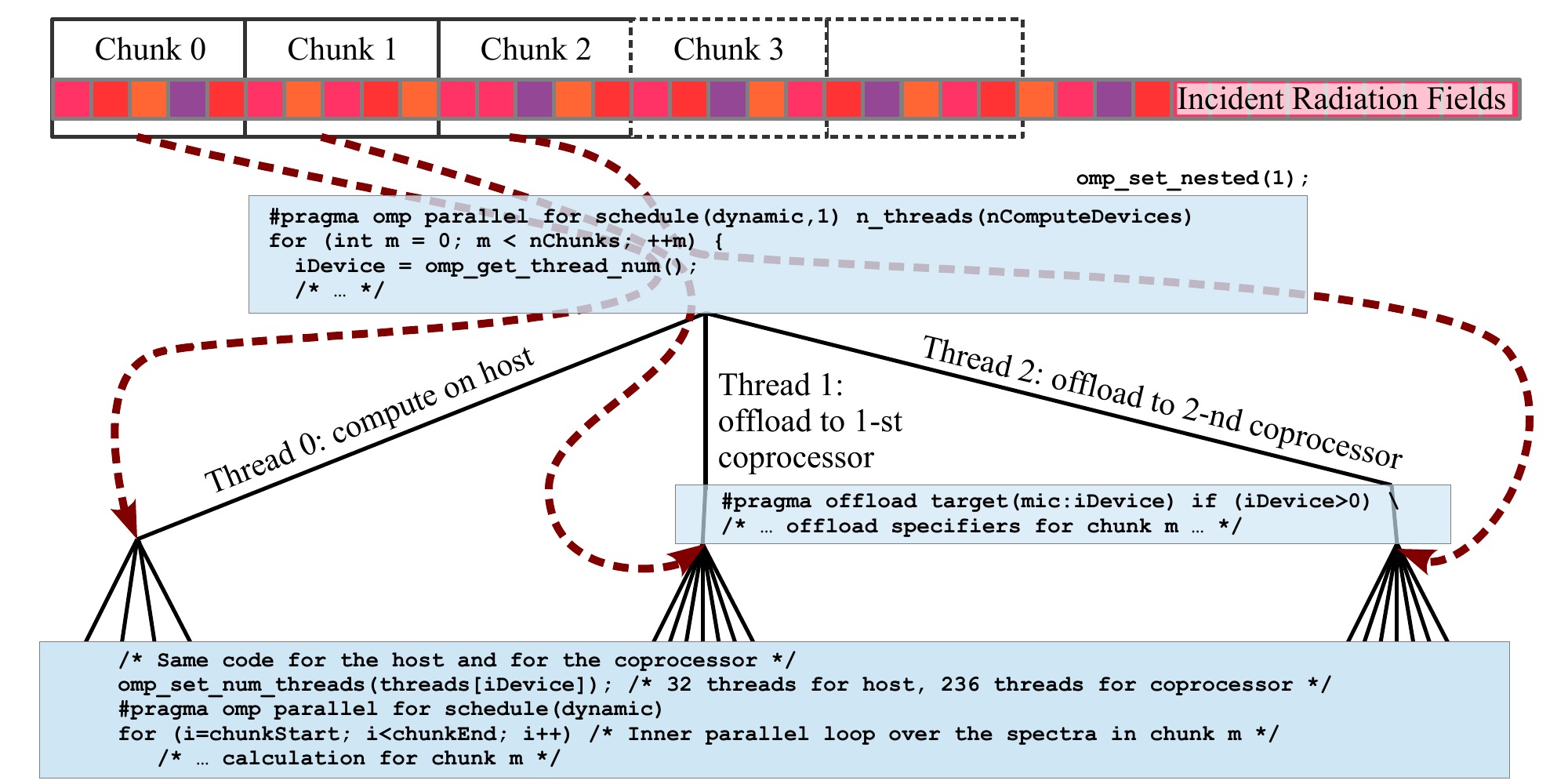}
\caption{Using nested parallelism and loop scheduling in OpenMP in order to dynamically schedule chunks of work on all available compute devices, including the host system and the coprocessors.\label{fig:heterogeneous}}
\end{figure*}

The utilization of multiple compute devices is essentially an additional level of parallelism.
By compute device in this discussion we mean either an Intel Xeon Phi coprocessor, or the host processor.
As with any task-parallel architecture, the scheduling of work across compute devices becomes important.
Scheduling in \lib\  must be dynamic, i.e., work must be distributed at runtime.
This is motivated by two factors:
\begin{enumerate}[1)]
\item The computing system is heterogeneous if the host and the coprocessor concurrently participate in calculations. 
In a heterogeneous system, work must be distributed proportionally to the relative performance of compute devices.
Otherwise some devices have to idle, awaiting the completion of work by other devices.
\item Different incident radiation spectra may take different amounts of time to process into stochastic emissivities, because the code uses pruning to exclude heating terms or whole spectra not interacting with grains in the stochastic heating mode (see \ref{sec:step1}).
\end{enumerate}

In order to utilize the additional level of parallelism (multiple compute devices), we use the blocking offload syntax and perform multiple concurrent offloads from the corresponding number of parallel threads on host. The loop scheduling functionality in OpenMP library takes care of scheduling.
When multiple compute devices are used to calculate the stochastic grain emissivity for \texttt{n} incident radiation fields, \lib\ splits these \texttt{n} tasks into multiple contiguous chunks (i.e., sets of radiation fields).
Then the application spawns a  thread-parallel region on the host with as many threads as there are compute devices.
Each thread in the parallel region is associated with one of the compute devices.
In computing systems enabled with Intel Xeon Phi coprocessors, the host system typically has no less than 8 logical cores (usually up to 32), but no more than 8 coprocessors (usually 1 or 2).
Therefore, we expect that all threads in this outer parallel region start simultaneously.
The OpenMP threads corresponding to compute devices are then used to run a parallel for-loop, which iterates over all chunks.
The loop scheduling mode is set to dynamic scheduling with a grain size of 1. 
In this mode, initially, one chunk from the pool is assigned to each thread, which then launches a blocking calculation for this chunk on the respective compute device.
When one of the threads (equivalently, one of the compute devices) completes the work and unblocks, another chunk from the pool is assigned to this thread (equivalently, to the corresponding compute device) by the OpenMP scheduler.
This process goes on until all chunks are processed.

The offloaded calculation spawns its own parallel loop, which distributes the calculations of all radiation fields in the current chunk across the cores of the device.
This parallel  loop uses as many threads as there are available logical cores on the respective compute device.

The host CPU can be used as a compute device, too, by mapping one of the threads to the host CPU.
In order to use the OpenMP loop that schedules across devices and, at the same time,  spawn OpenMP threads on the host processor for calculations, nested parallelism in OpenMP must be enabled.

The work scheduling technique described above is illustrated in Figure~\ref{fig:heterogeneous}.

\subsection{Runtime Environment}\label{sec:runtime}

In order to achieve maximum performance, some settings of the OpenMP and offload runtime library have to be set to non-default values.

Compute-bound workloads, such as the stochastic heating and emissivity calculation discussed here, usually benefit from binding OpenMP threads to physical or logical cores of the processor to prevent thread migration across cores.
For \lib, we recommend setting such binding using the OpenMP thread affinity interface.
On the host, prior to starting the calculation, the Linux environment variable \texttt{KMP\_AFFINITY} must be set by the user to the value ``\texttt{compact,granularity=fine}'' (hereafter, the quotes must not be included in the value of the environment variables).
This sets the OpenMP thread affinity on the host that binds OpenMP threads to the respective logical cores, and places threads with consecutive numbers as close to each other as possible.
Additionally, the environment variable \texttt{MIC\_ENV\_PREFIX} may be set to the value ``\texttt{MIC}'', the variable \texttt{MIC\_KMP\_AFFINITY} to the value ``\texttt{compact,granularity=fine}'', and \texttt{MIC\_PLACE\_THREADS} to ``\texttt{59C,4t}'' (for a 60-core coprocessor).
These settings effect the same type of OpenMP  affinity on all Intel Xeon Phi coprocessors in the system, ensuring that one core dedicated to OS-level offload tasks does not get loaded with computational work.

Another runtime optimization that improves the coprocessor performance is using large buffers for offloading data to coprocessors.
In order to use this feature with \lib, the environment variable \texttt{MIC\_USE\_2MB\_BUFFERS} on the host must be set to the value ``\texttt{64K}'', which automatically uses large buffers for all offloaded arrays exceeding 64~kilobytes in size.




\bibliographystyle{model1-num-names-cpc}
\bibliography{porter_vladimirov_transient_heating_cpc}







\end{document}